
\input psbox.tex
\psfordvips

\def\bbbc{{\mathchoice {\setbox0=\hbox{$\displaystyle\rm C$}\hbox{\hbox
to0pt{\kern0.4\wd0\vrule height0.9\ht0\hss}\box0}}
{\setbox0=\hbox{$\textstyle\rm C$}\hbox{\hbox
to0pt{\kern0.4\wd0\vrule height0.9\ht0\hss}\box0}}
{\setbox0=\hbox{$\scriptstyle\rm C$}\hbox{\hbox
to0pt{\kern0.4\wd0\vrule height0.9\ht0\hss}\box0}}
{\setbox0=\hbox{$\scriptscriptstyle\rm C$}\hbox{\hbox
to0pt{\kern0.4\wd0\vrule height0.9\ht0\hss}\box0}}}}

\def\build#1_#2^#3{\mathrel{\mathop{\kern 0pt#1}\limits_{#2}\limits^{#3}}}
\def\dyad#1#2{\mid #1\rangle\langle#2\mid }
\def\ket#1{{}\mid #1\rangle{}}

\def\braket#1#2{{}\langle#1\mid #2\rangle{}}
\def\obraket#1#2#3{{}\langle#1\mid #2\mid #3\rangle{}} 

\def\bbbr{{\rm I\!R}}

\font\eightrm=cmr8

\def\bk{{\bf k}}

\magnification=\magstep1

\centerline{\bf WEAK AND STRONG COUPLING REGIMES, VACUUM RABI SPLITTING} 
\centerline{\bf AND NONSTANDARD RESONANCES}

\vskip 0.8cm
\centerline{\bf C. Billionnet}
\centerline{\sevenrm Centre de Physique Th{\'e}orique, Ecole Polytechnique, 91128
  Palaiseau cedex, France}
\centerline{\sevenrm E-mail~: billionnet@cpht.polytechnique.fr}

\vskip 0.8 cm
\medskip
\underbar{Abstract.} For two discrete-level quantum systems in interaction, we follow the displacement in the complex plane of the eigen-energies of the compound system when the spectrum of one of the two systems becomes continuous. These new points are usually called resonances. This allows us to define and to calculate a critical value of the coupling constant which separates two well-known coupling regimes. We also give an example of these resonances for the hydrogen atom coupled to the continuum of states of the transverse electromagnetic field in the vacuum. We justify that some of the resonances be neglected.

with photons -\break  33.  Molecular spectra and interactions
with photons - 42.50.-p Quantum optics - 71.36.+c Polaritons -  71.38.-k Polaron and electron-phonon interactions - 73.21.-b Electron states and collective excitations in multilayers, quantum wells, mesoscopic, and nanoscale systems

\bigskip

\noindent

{\bf 1. Introduction}

\smallskip
In this work, we are interested in states in which a discrete-level quantum system ${\cal S}$ is coupled to a continuum ${\cal C}$. The total system may be an  atom coupled to the transverse electromagnetic field, in the vacuum or in a non perfect cavity, an electron in a quantum dot coupled to optical phonons or photons, an exciton coupled to optical phonons or photons in a microcavity. The continuum may also consist of electronic states, whereas ${\cal S}$ is a fixed energy photon.

Three points usually appear in the study of this question. The first one is the vacuum Rabi splitting, the fact that for a two level atom, for instance, the coupling of the atom to photons which are resonant with the transition splits the excited level. This is a simple fact of Quantum Mechanics (see for instance Cohen-Tannoudji {\it et al} 1973, p 408). The second point is the distinction between the weak and strong coupling regimes of the interaction of ${\cal S}$ and ${\cal C}$. The third one is the existence of bound states or almost bound states of the ${\cal S}$+ ${\cal C}$ system which occur at large coupling constant. We are going to show that these three points can be linked by the study of the resonances of the ${\cal S}$+ ${\cal C}$ Hamiltonian in the complex plane.

Many experimental studies have been performed in recent years as regards the second point, in the various domains we mentioned in the first paragraph. It is not possible to quote them all. Some of them are specially related to the transition between the two regimes. Let us mention for instance (Inoshita and Sakaki 97), where mixed electron-phonon states in a quantum dot (Reed 1993) are studied, or  (Verzelen {\it et al} 2000, Verzelen {\it et al} 2002, Hameau {\it et al} 1999, Boeuf {\it et al} 2000) and (Weisbuch {\it et al} 1992, Sermage {\it et al} 1996, Yamanoto {\it et al} 2000, Senellart 2003), about excitons coupled to phonons or photons, or also (Tignon {\it et al} 1995) in which the continuum is made of electronic states. Atoms in cavities are studied for instance in (Haroche 1984, Haroche 1992, Haroche et Raimond 1993, Berman 1994, Raimond {\it et al} 2001).
The resonances of the ${\cal S}$+ ${\cal C}$ system are numerous, as we showed it in (Billionnet 2004). This is in fact more or less known since a long time. But one usually considers two extreme situations. Either the imaginary parts of these resonances are practically zero (very narrow continuum) and the resonances  are very close to eigenvalues, thus easily identifiable, or some imaginary parts are very large, and the corresponding resonances are either not known, or deliberately ignored. The interest which has been taken recently in intermediate situations, such as those which occur in solid state physics for some electron-phonon couplings, leads to take all these resonances into account, without limiting oneself to perturbative calculations.

In these intermediate situations, the coupling constant and the continuum's width  have various values. The continuum's width actually depends on the states of ${\cal S}$. As regards the interaction of an atom with the transverse electromagnetic field, the coupling constant in the interaction Hamiltonian is indeed the fine structure constant, but the details of the effective coupling, its dependence with respect to the energy of the photon emitted in some transition, depends on the transition which is considered. For instance, in general, the spatial extension of the atom's states is a factor which affects the width of the continuum as it is seen by the atom. The larger the spatial extension, the narrower the continuum's shape and the closer to the reals the resonances. This influence of the spatial extension may be very important (Billionnet 2001). We shall use the coupling to the photon of the hydrogen atom Rydberg states to illustrate this fact again and also to prepare possible later studies of large molecules. In the former case, we will show that 
only resonances perturbed from the free atom's energies are of interest, although states with principal quantum number $n$ extend over a distance proportional to $n$. Wavelengths of transitions between two states are too large with respect to the mean extension of these states for other resonances to be of interest. It is nevertheless a fact that these other resonances exist and we will calculate one of them.  

In section 2, we first introduce the question in a general and qualitative way. Then, through a two level model, we study the behaviour of the resonances under the variation of three parameters: the coupling constant, the continuum's width and a continuum/system detuning. This will lead us to a precise definition of a transition point between the strong and weak coupling regimes. We will meet the V.R.S. in the narrow continuum limit (strong coupling regime). In section 3, we will apply the general preceding analysis to several situations among those mentioned in the beginning of this introduction. Section 4 is devoted to the hydrogen atom.

\medskip
{\bf 2. General study of a discrete-level system coupled to a continuum}
\smallskip

It is now clear that for any discrete-level system ${\cal S}$ coupled to a massless field, or more generally to a continuum, the number of eigenvalues or resonances of the Hamiltonian is much greater than the number of levels of ${\cal S}$. The term resonances here means poles of matrix elements of the Hamiltonian's resolvent. Even with photons having only one possible state, if their number is not limited, the number of these eigenvalues or resonances is already infinite. Now, the number of linearly independent possible states for each photon may be infinite and the number of discrete levels is itself infinite. This make three reasons why the Hamiltonian operates in an infinite dimensional space. In the case where ${\cal S}$ is coupled to a massless field, let us denote the Hilbert space of ${\cal S}$ by ${\cal H}_S$ and that of the field by ${\cal H}_{\rm rad}$. In a $N$-dimensional space, a hermitian matrix has $N$ real eigenvalues (possibly degenerated). In (Billionnet 2005), we showed that the number of resonances is comparable to the dimension of the states of the total system ${\cal S}$+ field rather than to the number of discrete ${\cal S}$-states. Since these two numbers are infinite, (more precisely card $(I\!\!N))$, we have to make this statement more precise: for any restriction of the Hamiltonian to finite-dimensional subspaces of ${\cal H}_{\cal S}\otimes{\cal H}_{\rm rad}$, the number (with degeneracies taken into account) of resonances of the restricted Hamiltonian is the dimension of these subspaces. (This dimension is not necessarily the product of the dimension of a subspace of ${\cal H}_{\cal S}$ by the dimension of a subspace of ${\cal H}_{\rm rad}$.)

\medskip
{\bf 2.1. Standard and nonstandard resonances}
\smallskip

When the continuum's width is large, some of these resonances are the familiar ones which appear through the perturbative approach to the coupling: they are the energies of the ${\cal S}$-levels moved into the complex plane by the coupling. We call them standard resonances, according to the general following definition:

\underbar{\sl Definition {\rm 1}}: {\it In the coupling of a discrete-level system ${\cal S}$ to a continuum ${\cal C}$, standard resonances (or eigenvalues) are resonances (or eigenvalues) which tend to the energies of eigenstates of ${\cal S}$, when the coupling constant $\lambda$ tends to $0$, the Hamiltonian being of the form $\lambda V$. We call resonances (or eigenvalues) which do not have this property nonstandard resonances (or eigenvalues).}

Let us note that ${\cal S}$ and ${\cal C}$ do not play a symmetrical role in this labelling. Let us also note that ${\cal S}$ may be a material system and the continuum the set of states of the radiation. But it may be the other way round (see section 3.2). We refer to the next to last paragraph of section 2.3 for a justification of the term ``nonstandard''.

Of course, the study of these resonances requires that a Hamiltonian be given, but we will begin with general considerations, without specifying the interaction. In the whole section 2, the continuum is that of the states of a free scalar photon.

\medskip
{\bf 2.2. Coupling functions}
\smallskip

Let us denote the eigenstates of ${\cal S}$ by $\ket 0,\ket
1,\cdots$ and set
$$
H=\sum_n E_n\dyad nn\otimes 1+1\otimes H_{\rm rad}+H_I    \eqno (1)
$$
the Hamiltonian of the ${\cal S}+field$ system. $H_{\rm rad}$ is the Hamiltonian of the free field.
Let us assume that for all $n>m$ and all $\varphi\in {\cal
  F}_1$, the one-photon-state space, there are functions $g_{nm}$ such that
$$
\obraket n {H_I} {m;\varphi}=\int\varphi(k)\ g_{nm}(k)\ dk   \eqno (2)
$$
(we set $\ket{m;\varphi}=\ket m\otimes\varphi$). Thus, formally, we have
 
$$
g_{nm}(k)=\obraket n {H_I} {m;k}                     \eqno (2')
$$
and $g_{nm}(k)$ describes the coupling of state $\ket m$ to state $\ket n$ through absorption of a photon with wave-vector $k$. We call $||g_{nm}||_2^{-1}\ g_{nm}$ the coupling function. ($||g_{nm}||_2$, the $L^2$-norm, has the dimension of an energy.) Generally, $H_I$ has a lot of other a priori non-zero matrix elements than (2). We consider the following approximation of $H$ 
$$
H^{\rm app}:=\sum_n E_n\dyad nn\otimes 1+1\otimes H_{\rm
  rad}+H_I^{\rm app}                             \eqno (3)
$$
with
$$
H_I^{\rm app}:=\sum_{m<n}\Big(\dyad nm\otimes\  a(g_{nm})+\dyad mn\otimes\ 
\big(a(g_{nm})\big)^*\Big)                     \eqno (3')      
$$
where $a(.)$ is the field's annihilation operator. This Hamiltonian neglects matrix elements of $H_I$ between states the number of photons of which differs by more than one, as well as matrix elements $\obraket m {H_I} {n;k}$ with $m<n$.

\medskip

{\bf  2.3. An example of a couple of a standard and a nonstandard resonance: the vacuum Rabi splitting}

\smallskip
Nonstandard resonances are easily seen in the limit where each $g_{nm}$ gets peaked around a value $k_{nm}$. Let us show this. The interaction Hamiltonian (3) becomes
$$
H_I^{\rm dis}:=\sum_{m<n}\lambda_{nm}\Big(\dyad nm\otimes\  a_{k_{nm}}+\dyad mn\otimes\ 
a_{k_{nm}}^*\Big)  \ .                          \eqno (4)
$$
Let us set
$$
H^{\rm dis}:=\sum_n E_n\dyad nn\otimes 1+1\otimes \sum_{m<n} \hbar c k_{nm}\ a^*_{k_{nm}}a^{}_{k_{nm}} +
H_I^{\rm dis}        \ .                                    \eqno (5)
$$
Eigenvalues of $H^{\rm dis}-H_I^{\rm dis}$ are $E_n$, $E_n+\hbar c k_{ij}$, 
$E_n+\hbar c k_{ij}+\hbar c k_{lm}$, etc. . When the $\lambda_{nm}$'s are small, one can expect the eigenvalues of $H^{\rm dis}$ to be close to the preceding ones. Then, except for particular values of the $k_{ij}$'s, only the eigenvalues of $H^{\rm dis}$ which are close to the $E_n$'s are standard, in the sense of definition 1. Others are nonstandard: when the coupling constant goes to $0$, they tend to a linear combination of an atomic level's energy (possibly with the coefficient $0$) with energies of a non-zero number of photons. In the particular two-level case, the excitation number operator $N:=\sum_{n=0,1}\dyad nn\otimes 1+1\otimes a^*_{k_{01}}a^{}_{k_{01}}$ commutes with $H^{\rm dis}$. Let ${\cal E}_l$ be the eigenspace associated with eigenvalue $l$ of $N$. For $l\geq 1$, its dimension is $2$. For example, for $l=1$, ${\cal E}_1$ is spanned by $\ket {0,k_{01}}$ and $\ket{1,\Omega}$, $\Omega$ denoting the vacuum in the radiation space. In this case, the restriction of $H^{\rm dis}$ to ${\cal E}_1$ has two eigenvalues which are close to $E_1$ and $E_0+\hbar c k_{01}$ respectively. We called the former standard and the latter nonstandard.
If the photon energy $k_{01}$ equals $E_1-E_0$ (resonance), then the coupling $H_I^{\rm dis}$ removes the degeneracy of the eigenvalue of $H^{\rm dis}-H_I^{\rm dis}$ associated with eigenvectors in ${\cal E}_1$, and this is called the vacuum Rabi splitting. For this special value of the photon's energy, the standard and nonstandard eigenvalues turn into the doublet of the V.R.S.

The number of eigenvalues of $H^{\rm dis}$ is infinite. Note that even the number of standard eigenvalues is greater than the number of discrete states, since the displacements of $E_1$ calculated in the ${\cal E}_l$'s are a priori different. This is also true for resonances in the non-zero width case. We know (see (Billionnet 2005), for instance) that for two levels, with the coupling function $g(p)\sim p/(1+p^2)$, there is a standard resonance and a nonstandard one for the restriction of Hamiltonian (3) to each ${\cal E}_l$.  

When the width of the $g_{nm}$'s is not zero, we expect the eigenvalues either to remain eigenvalues or to become resonances. If the detuning is large, that is to say if the photons' energy is not resonant with any atomic transition, and if the coupling constant is small, then resonances which were nonstandard at zero width will remain so if the width is sufficiently small. For example, for $r\not=l$, eigenvalues close to $E_r+(l-r)k_{ij}$ will remain nonstandard resonances.

In this setting, one may not see any reason for such a term, for resonances which are simply perturbed values of eigenvalues, and thus have a simple physical meaning. The labelling has been introduced in a case where these resonances exist but are not obvious, the case of an atom in the vacuum. They are very different from the atomic levels and we need a term to label them. The term is kept in other cases.

We are now going to study the two-level case more thoroughly, so as to show how  useful it is to pay attention to all resonances. We are going to vary different parameters of the coupling of ${\cal S}$ to ${\cal C}$ and to follow the trajectories of some resonances under these variations.

\medskip
{\bf 2.4. Study of a two level system}
\smallskip

We consider a two-level system in the RWA approximation. So there is only one coupling function $g$. Let $E_1>0$ be the energy of the excited state $\ket 1$ and let us assume that the energy of the fundamental state $\ket 0$ is $0$.
The Hamiltonian is
$$
H=E_1\dyad 11 \otimes\  1+1\otimes H_{\rm rad}+\lambda\ \Big(\dyad
01\otimes\  \big(a(g)\big)^*+\dyad 10\otimes\  a(g)  \Big) \ .  \eqno (6)
$$
We assume that $||g||_2=1$, the strength of the coupling appearing in $\lambda$, which has the dimension of an energy.

\noindent
As the resonances can only be obtained by computer, we are going to chose a particular $g$. This example will yield the important notions. The chosen function is
$$
g(k):=\sqrt{2\over\pi}\ {(\mu k_0)^{-{1/2}}\over 1+\mu^{-2}({k\over
    k_0}-1)^2} \ .                                   \eqno (7)
$$
As $\mu$ gets smaller, the function becomes more peaked at $k_0>0$
(the width is $2\mu k_0$, see figure 2). We set $\delta:=E_1/ (\hbar ck_0)-1$; $\delta\in]-1,\infty[$. When $g$ is very peaked at $k_0$, $\delta$ measures the detuning between the levels' spacing and the energy of the coupled photons. We will only consider eigenvalues or resonances of the restriction of $H$ to ${\cal E}_1$ (defined in section 2.3). They are zeros of 
$$
z\rightarrow z-E_1-\lambda^2\int_{-\infty}^{+\infty} {g(k)^2\over z-\hbar ck}\ dk\eqno (8)
$$
or of its analytic continuation into the lower complex half-plane, across the cut $\bbbr^+$. (For $z=E_1$, the integral term in (8) is simply the correction to the upper level's energy due to the emission and re-absorption of a virtual photon.)
With $\kappa=(\hbar ck_0)^{-1}\lambda$ and for $\zeta<0$, let us set
$$
f(\kappa,\mu,\delta,\zeta):=\zeta-(1+\delta)-{2\kappa^2\over\pi\mu}\int_{-\infty}^{+\infty} {1\over \big(1+({y-1\over
  \mu})^2\big)^2}\ {1\over \zeta-|y|}\ dy      \ .           \eqno (8')
$$
The resonances we are interested in are the product of $\hbar ck_0$ and zeros of
the analytic continuation $f_+(\kappa,\mu,\delta,.)$ of $f(\kappa,\mu,\delta,.)$
 into the lower complex half-plane.

\noindent
We are now going to study the position of these zeros as functions of three physical parameters of the system:  $\kappa,\ \mu$ and $\delta$. An important point has to be mentioned: when at least two variables are considered, the position of these zeros is a multivalued function of these variables, even if these variables remain real (Billionnet 2002). This leads to a complication as regards the zeros' notation. 

\smallskip
{\bf 2.4.1. Displacement of two resonances through the variation of the coupling function's width}

We start with a study with clearly non-zero detuning in order to study the effect of the variation of the continuum's width independently of resonance effects (here this word means zero detuning).

When $\mu$ tends to $0$, it can be shown that $f(\kappa,\mu,\delta,\zeta)$ and $f_+(\kappa,\mu,\delta,\zeta)$ tends to
$\zeta-1-\delta-\kappa^2/( \zeta-1)$. For small $\kappa$, one of the zeros is close to $1+\delta$ (resonance close to $E_1$) and the other one is close to $1$ (resonance close to $E_0+\hbar ck_0$). We denote the former by $\zeta_{w,at}$ and the latter by $\zeta_{w,ph}$. The index $w$ indicates that only the width varies; $\delta$ and $\kappa$ remain constant. Subscripts "at" and "ph" indicate that the $\mu\rightarrow 0$ limits are the energy of the atom's excited state and the energy of the photon, respectively. Let us mention that $f_+(\kappa,\mu,\delta,.)$ has another zero whose physical meaning is no more obvious. It is described in Appendix A.

\noindent
For $\kappa=0.1$ and $\delta=0.25$, the position of two resonances when $\mu$ is varied is given by the curves of figure 1.


\vskip 0.4cm

\setbox11=\hbox{\psannotate{\psboxto(5.8cm;0cm){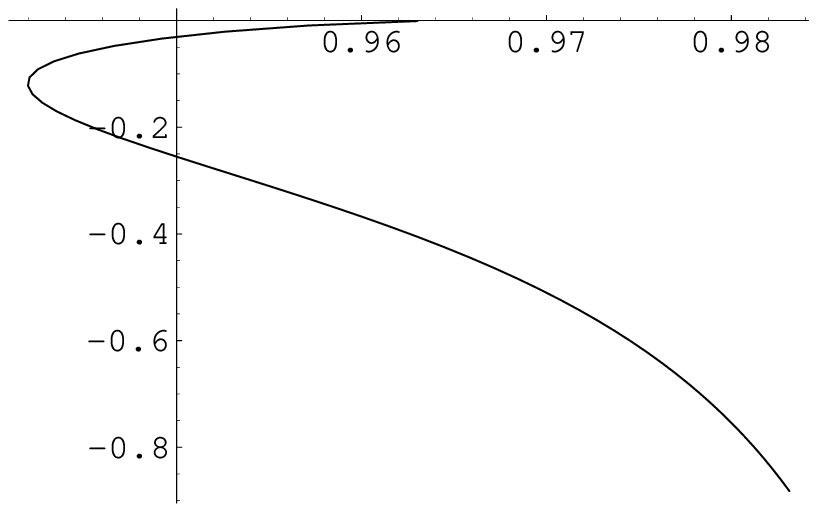}}
{\at{4cm}{0.3cm}{$\mu=1\ _{\searrow}$}\at{2.8cm}{1.5cm}{$\zeta_{w,ph}$}}}

\setbox121= \vbox{   \boxit {  \vbox{\vskip0.2cm\hbox{ $\ \kappa=0.1\
        \ $} \hbox{
      $\ \delta=0.25\ \ $}\vskip0.2cm}}    \vskip 1.5cm}

\setbox13=\hbox{\psannotate{\psboxto(5.8cm;3.3cm){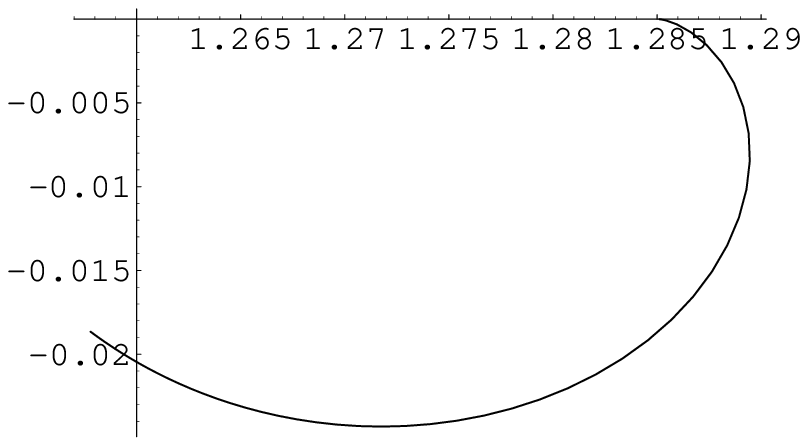}}{\at{0.7cm}{0.9cm}{$\leftarrow\mu=1$
    }\at{3.3cm}{0.8cm}{$\zeta_{w,at}$}}}

\hbox{\raise -4mm\box11\box121\box13}

Figure 1. Values $\zeta=(\hbar ck_0)^{-1}\ z$, for two resonances $z$,
$\mu$ varying from $0.01$ to $1$
\vskip 0.5cm


\noindent
Limits of $\zeta_{w,ph}(\mu)$ and $\zeta_{w,at}(\mu)$ for $\mu$ tending to $0$ are respectively
$1+2^{-1}(\delta-(\delta^2+4\kappa^2)^{1/2})=0.965$ and $1+2^{-1}(\delta+(\delta^2+4\kappa^2)^{1/2})=1.285$.

\noindent
The coupling functions for $\mu=0.01$
and $\mu=1$ are plotted in figure 2.


\vskip 0.4cm

\setbox21=\hbox{\psannotate{\psboxto(5.8cm;3.3cm){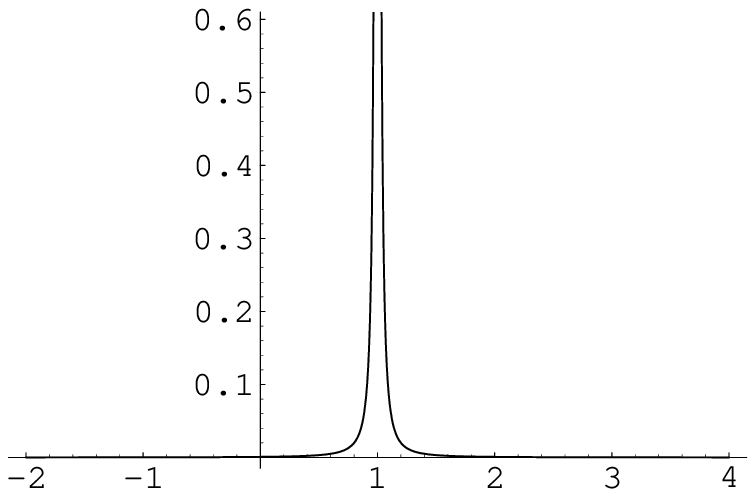}}{\at{3cm}{3cm}
{$g(1)\simeq 8$}}}

\setbox22=\hbox to 3cm{}

\setbox23=\psboxto(5.8cm;0cm){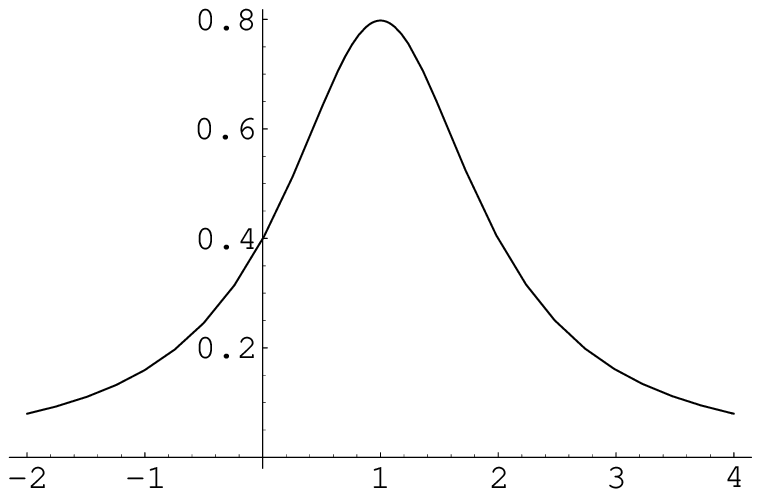}

\hbox{\box21\box22\box23}

Figure 2. The coupling function for $\mu=0.01$ (left) and
for $\mu=1$ (right). 

\hskip 1.5cm Units are $k_0$ on the abscissa and $k_0^{-1/2}$ on the ordinate.

\vskip 0.5cm


In figure 1 we see that the imaginary part of $\zeta_{w, at}$ does not exceed $0.03$ in modulus whereas that of $\zeta_{w, ph}$ increases and takes much larger values when $\mu$ increases.

For small $\mu$, the resonances are close to the real axis. However, let us note that $\zeta_{w,at}$ is here also much closer to the reals than $\zeta_{w,ph}$. Indeed, the calculation gives  $\zeta_{w,at}=1.285-2.7\times  10^{-6}\ i$ and $\zeta_{w,ph}=0.963-9.8 \times 10^{-4}\ i$ for $\mu=0.01$.

Before we turn to the $\mu\rightarrow 0$ limit, let us comment on the standard or non-standard character of $\zeta_{w,at}(\mu)$
and $\zeta_{w,ph}(\mu)$, with the same values of $\kappa$ and $\delta$. Since, by a continuity argument, $\zeta_{w,ph}$ and $\zeta_{w,at}$ remain respectively in neighbourhoods of $0.965$ and $1.285$ when $\mu$ is small, $\zeta_{w,at}(\mu)$ is standard for small $\mu$, whereas $\zeta_{w,ph}(\mu)$ is nonstandard. Indeed, the energy of $\ket 1$ is $1.25$, in $\hbar ck_0$ units, and it is actually the zero of $f$ sitting at $\zeta_{w,at}(\mu)$ for $\kappa=0.1$ which tends to 1.25 when $\kappa$ tends to $0$. In section 2.4.3.2, we show what happens when $\mu$ increases. Physically, when $\mu$ gets sufficiently large, the detuning is no longer noticeable and, if the coupling is strong enough, we may expect that the atomic and photonic states be mixed, and even hardly distinguishable. As a consequence, if $\mu$ is large, it is difficult to guess which of the two resonances goes to $1$ and which goes to $1+\delta$, when $\kappa$ goes to $0$. In other words it is difficult to guess which is the standard one. In section 2.4.3.2, we even show that for some value of $\kappa$ and $\mu$, both resonances coincide. 

In the $\mu\rightarrow 0$ limit, the Hamiltonian formally becomes
$$
H_1=E_1\dyad 11 \otimes\  1+1\otimes H_{\rm rad}+\lambda\ (\dyad
01\otimes\  a_1^*+\dyad 10\otimes\  a_1 )   \eqno (9)
$$
where $a_1$ is the annihilator of a photon with energy $\hbar ck_0$. Photons whose wave numbers differ from $k_0$ are decoupled. Let us consider the reduced Hilbert space ${\cal H}_0$, tensor product of ${\cal H}_S$ and the $k_0$-photon's Fock space. The restriction of $H_1$ to ${\cal H}_0$ has an infinite number of eigenvalues. They are of the form $z_{\pm,n}=\hbar ck_0\ \zeta_{\pm,n}$, with
$$
\zeta_{-,n}=n+2^{-1}(\delta-\sqrt{\delta^2+4n\kappa^2})\ ,\quad
\zeta_{+,n}=n+2^{-1}(\delta+\sqrt{\delta^2+4n\kappa^2})\ .  \eqno (10a)
$$
The associated eigenvectors are
$$
\phi_{\pm,n}=\big(1+n\kappa^2\zeta_{\pm,n}^{-2}(\kappa)\big)^{-1}\Big(\ket
1\otimes\ket{k_0}^{\otimes n}+\sqrt n\  \kappa\ \zeta_{\pm,n}^{-1}(\kappa)\ket
0\otimes\ket{k_0}^{\otimes (n-1)}\Big)   \ .             \eqno (10b)
$$
The coupling thus yields mixed states. When $\kappa$ goes to
$0$, $\phi_{+,n}$ tends to $\ket
0\otimes\ket{k_0}^{\otimes n}$ for $\delta>0$ and to
$1\otimes\ket{k_0}^{\otimes (n-1)}$ for $\delta<0$. It is the other way round for $\phi_{-,n}$.

These mixed states do not exist anymore as eigenstates of the Hamiltonian when the coupling function has a certain width. The eigenvalues, i.e. the energies of these states, turn into the resonances drawn in figure 1. They both acquire an imaginary part.

In the non-zero width case, let us now look at what happens when the two levels' spacing is varied around $\hbar ck_0$.

\smallskip
{\bf 2.4.2. Variation with respect to the detuning. The levels' anti-crossing.}

a) {\it The discrete case.}
Let us first recall what happens in the case where the width of $g$ is zero. When $\delta$ varies, both energies (10a) of the mixed states corresponding to $n=1$ vary. When $\delta$ tends to $\pm\infty$, they asymptotically tend to the energies of states $\ket 0\otimes\ket{k_0}$ and $1\otimes\ket 0$. These limits (in $\hbar ck_0$ units) are drawn in dashed lines in figure 3, for $\kappa=0.1$. They cross when $E_1=\hbar ck_0$.


$$
\psannotate{\psboxto(6cm;0cm){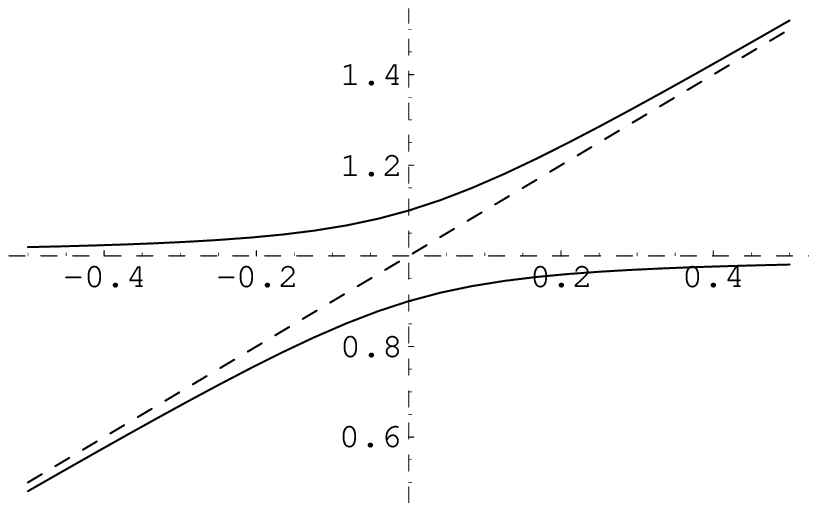}}{\at{6.1cm}{1.8cm}{$\delta$}\at{2.7cm}{3.9cm}{$E/\hbar ck_0$}\at{4cm}{1.4cm}{$\zeta_-$}\at{2cm}{2.3cm}{$\zeta_+$}}
$$
\centerline{Figure 3. The levels' anti-crossing, for an infinitely narrow continuum.}

\smallskip


\noindent
We see that the interaction yields what is called an anti-crossing, for whatever value of the coupling constant. The larger the coupling constant, the larger the repulsion of the two curves, since the energies at $\delta=0$ are separated by $2\kappa$. The same phenomenon repeats in the neighbourhood of $n$-photon resonances. We are now going to show how this anti-crossing is modified when the width of $g$ is no longer zero.

b) {\it The narrow continuum case.}
To each point on one of the two curves of the discrete case, there now corresponds, in the continuous case, a complex number. For example, for $\mu=0.01$ and $\delta=0.25$, the curves in figure 1 give values $0.963-9.8\times 10^{-4}$ and
$1.285-2.7\times 10^{-6}$. When $\delta$ varies, with the same $\mu$ value, the resonances move in the lower complex half-plane as is indicated in figure
4.

In this figure, we see that the imaginary parts of both resonances are more or less the same for $\delta=0$, about $-9.5\times 10^{-5}$. The one whose real part is greater than $1$ will be denoted by $\zeta_+$, the other by $\zeta_-$. For both curves, it can be shown that the imaginary part tends to $-\mu=-0.01$ when the real part goes to $1$. (We recall that $1-i\mu$ is a pole of the integrand in (8'), coming from a pole of $g$).

We also see that $\zeta_+$ is asymptotic to the reals when $\delta\rightarrow +\infty$. Conversely, $\zeta_-$ comes closer to the reals when $\delta$ decreases to $-1$, and tends to $1-i\mu$ when $\delta\rightarrow +\infty$. One usually considers the imaginary part of the resonance associated with an excited atomic state as the energy half-width of this state, a state that the coupling has made unstable. In the same way here, we may say, as in the discrete case, that the resonance $\zeta_-$ tends to the photon's energy in the limit $\delta\rightarrow +\infty$, it being understood that this energy is spread over a width equal to $2\mu$. 

\vskip1cm

\setbox41=\hbox{\psannotate{\psboxto(6cm;0cm){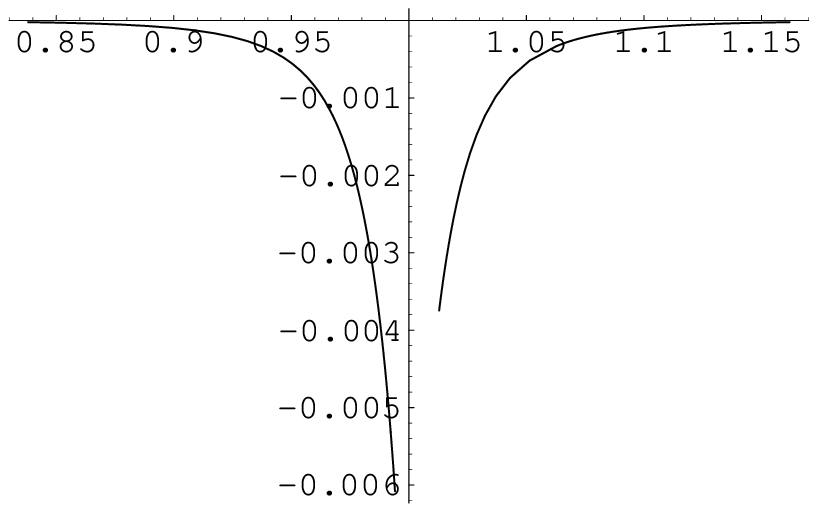}}{\at{6.1cm}{3.1cm}{$\nwarrow
    \delta=0.1$}\at{3.3cm}{1.4cm}{$\leftarrow
    \delta=-1$}\at{1.7cm}{-0.3cm}{$\delta=3\nearrow$}\at{0cm}{3cm}{$\uparrow$}\at{0cm}{2.4cm}{$\delta=- 0.1$}\at{1.3cm}{2cm}{$\zeta_-$}\at{4cm}{2cm}{$\zeta_+$}}}
\setbox42=\hbox to 3cm{}

\setbox43=\hbox{\psannotate{\psboxto(5cm;0cm){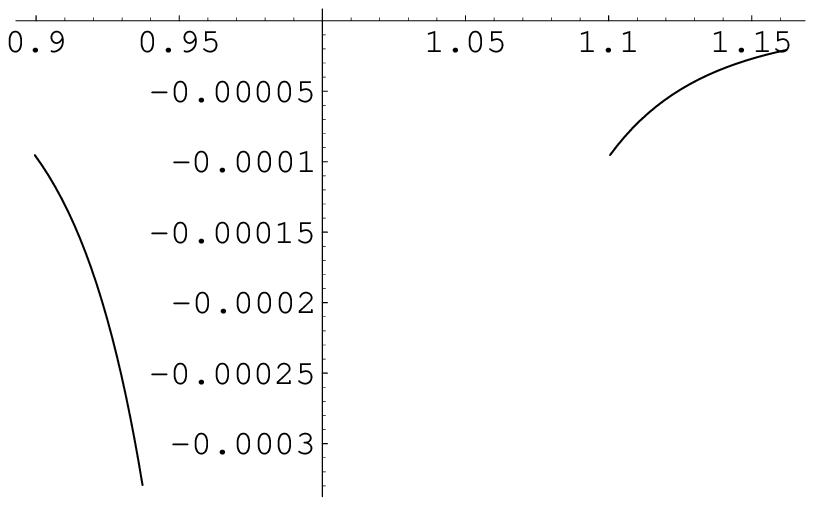}}{\at{-1.2cm}{1.9cm}{$\delta=0\nearrow$}\at{-0.8cm}{0.1cm}{$\delta=0.1\rightarrow$}\at{4cm}{3.5cm}{$\delta=0.1$}\at{4.7cm}{3.2cm}{$\downarrow$}\at{2.2cm}{2cm}{$\delta=0\rightarrow$}\at{0cm}{1cm}{$\zeta_-$}\at{4.2cm}{2.3cm}{$\zeta_+$}\at{1cm}{-0.7cm}{{\eightrm
      enlargement}}}}

\hbox{\box41\box42\raise 0.57cm\box43}

\hskip 2cm

\noindent
Figure 4. Variation of two resonances in the complex plane, with respect to the detuning

\hskip 0.8cm $\delta$, for $\kappa=0.1$ and $\mu=0.01$ (expressed in $\hbar ck_0$ units).

\smallskip

This leads us to propose to represent the mixed states' energies in the following way, which generalizes the diagram in figure 3. 

$$
\psboxto (6cm;0cm){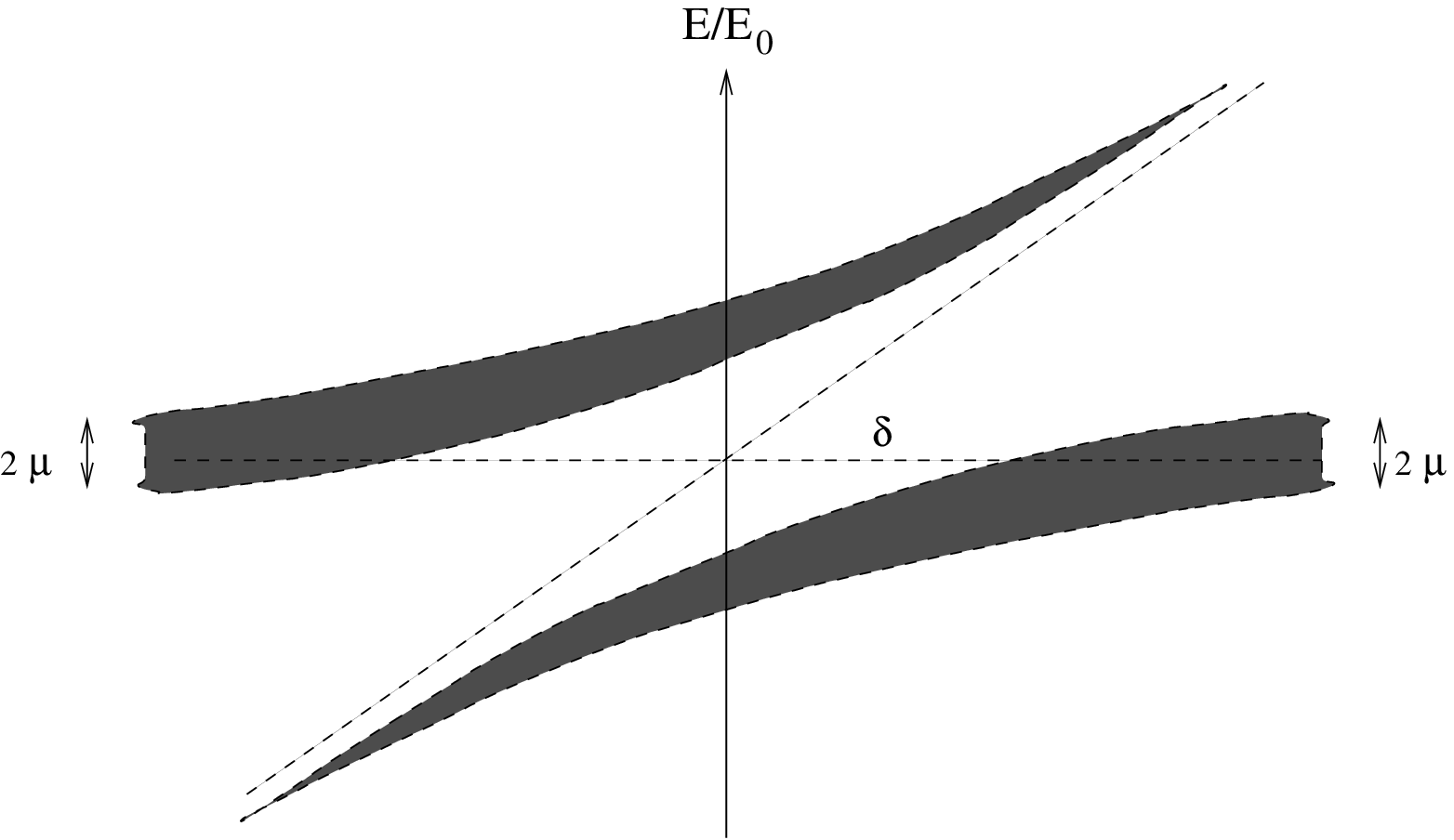}
$$
Figure 5. Representation of two resonances in the real plane, as a function of the detuning, 

\hskip 0.8cm for $\kappa=0.1$ and $\mu=0.01$.

\smallskip


For each value of $\delta$, a resonance is represented by a vertical line segment centered at the real part of the resonance and whose length is twice the imaginary part, i.e. the set $\delta+i[\Re \zeta-\Im\zeta,\Re \zeta+\Im\zeta]$.

c) {\it Discretization of the continuous case.} In some papers, things are presented in an other way. The continuum is discretized into a set of photon wave-vectors $k_1,\cdots,k_n$, with corresponding coupling constants $\lambda_1,\cdots,\lambda_n$. In the one-excitation space ${\cal E}_1$, we thus get an Hamiltonian whose only non-vanishing matrix elements are those between states $\ket{1;0}$ and
$\ket{0;k_i}$. As an example, let us take three values for $k$, say $k_-:=k_0(1-\mu),\ k_0,\ k_+:=k_0(1+\mu)$ and
coupling constants $\lambda_-:={1\over 2}\kappa\  \hbar ck_0,\ \lambda_0:=\kappa\  \hbar ck_0,\
\lambda_+:={1\over 2}\kappa\  \hbar ck_0$. Then, in the $\{ \ket{1;0},
\ket{0;k_-},\ket{0;k_0},\ket{0;k_+}\}$ basis, the Hamiltonian's matrix is
$$M=\hbar ck_0\pmatrix{1+\delta&\kappa/2&\kappa&\kappa/2\cr
\kappa/2&1-\mu&0&0\cr
\kappa&0&1&0\cr
\kappa/2&0&0&1+\mu}
$$
Let us set $\kappa=0.1$ and $\mu=0.01$, as before. The variation with $\delta$ of the four eigenvalues of $M$ is given in figure 6.


\setbox61=\hbox{\psannotate{\psboxto(5cm;0cm){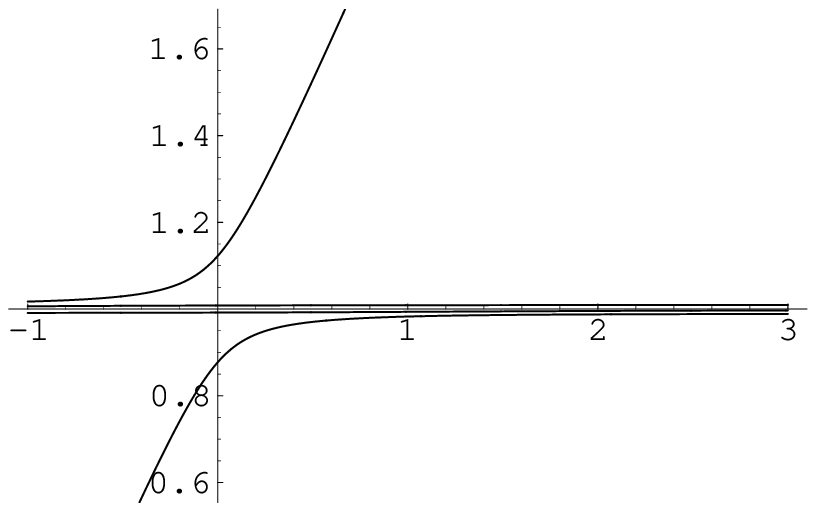}}{\at{0.5cm}{0.2cm}{$\zeta_1$}\at{2.1cm}{2.7cm}{$\zeta_4$}\at{5cm}{1cm}{$\delta$}\at{0.8cm}{3.3cm}{$E/\hbar ck_0$}\at{2.1cm}{1.4cm}{$\swarrow$}\at{2.1cm}{1.7cm}{$\zeta_2$
    and $\zeta_3$}}}
\setbox62=\hbox to 2cm {}
\setbox63=\hbox{\psannotate{\psboxto(5cm;0cm){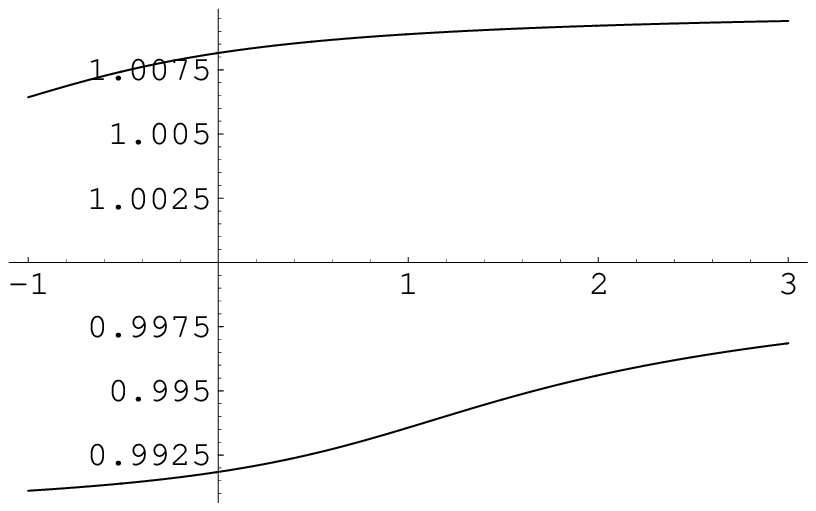}}{\at{3cm}{0.3cm}{$\zeta_3$}\at{3.3cm}{3.2cm}{$\zeta_2$}\at{5.1cm}{1.4cm}{$\delta$}\at{0.8cm}{3.3cm}{$E/\hbar ck_0$}}}

$$\hbox{\box61\box62\box63}$$
Figure 6. Four eigenvalues of the Hamiltonian, when the
continuum is replaced by 

\hskip 0.85cm three discrete values (in $\hbar ck_0$ units). $\kappa=0.1$.

\medskip


It can be shown that the four curves do not cross. Therefore, it is $\hbar ck_+$, the greatest of the three eigenvalues in the continuum for $\delta\rightarrow -\infty$, which tends to the energy of state  $\ket 1$ when $\delta\rightarrow
+\infty$. Conversely, it is $\hbar ck_-$, the smallest of the three eigenvalues in the continuum for $\delta\rightarrow +\infty$, which tends to the energy of state  $\ket 1$ when $\delta\rightarrow -\infty$. The description we gave in the continuous case is a concise rigorous way of conveying what may be approached by such discretizations.

\noindent

Looking back at figure 5, a visualization of figure 4, we see that we get states which are not only mixed states, but also, in a sense, enlarged states. Let us now look at the dependence with respect to $\lambda$. It will enable us to examine the notions of strong and weak coupling regimes in the light of the preceding results.

\smallskip
{\bf 2.4.3. Variation with the coupling constant} 

2.4.3.1. {\it A change in the regime around a critical value.} Let us qualitatively see what is expected. When $\kappa$ decreases, $\mu$ being fixed, the two grey tinted regions of figure 5 approach each other. Since their widths at each end do not depend on $\kappa$, the line segments at $\delta=0$ are going to overlap. The situation is then that of figure 7a. The calculation 

\setbox71=\hbox{\psannotate{\psboxto(6cm;0cm){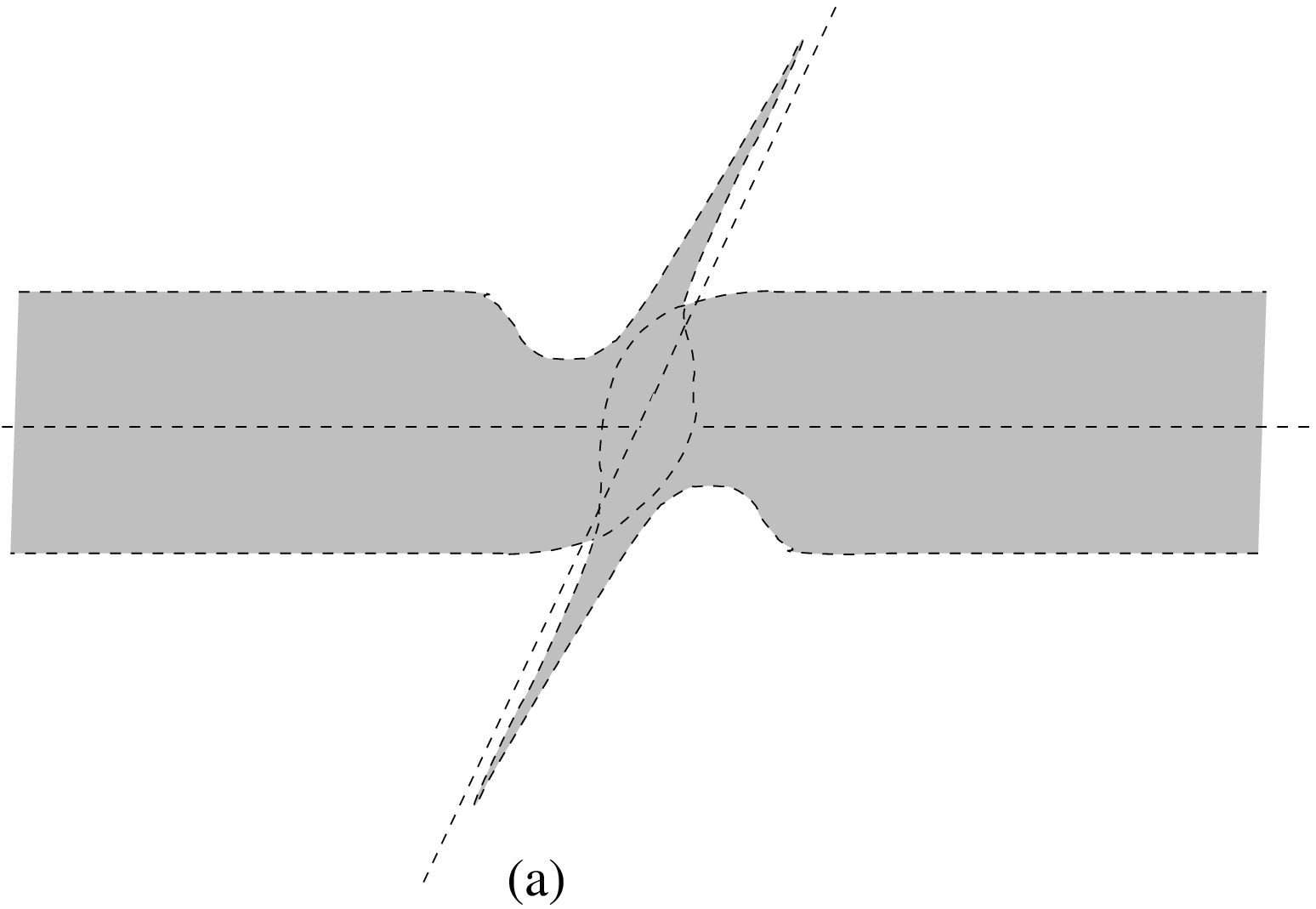}}{\at{6cm}{2cm}{$\delta$}\at{1.5cm}{3.4cm}{\boxit{\hbox{$\
          \kappa>\kappa_c\ $}}}\at{1cm}{1.2cm}{$\zeta_+$}\at{4cm}{1.2cm}{$\zeta_-$}}}
\setbox72=\hbox to 1cm{}
\setbox73=\hbox{\psannotate{\psboxto(5cm;0cm){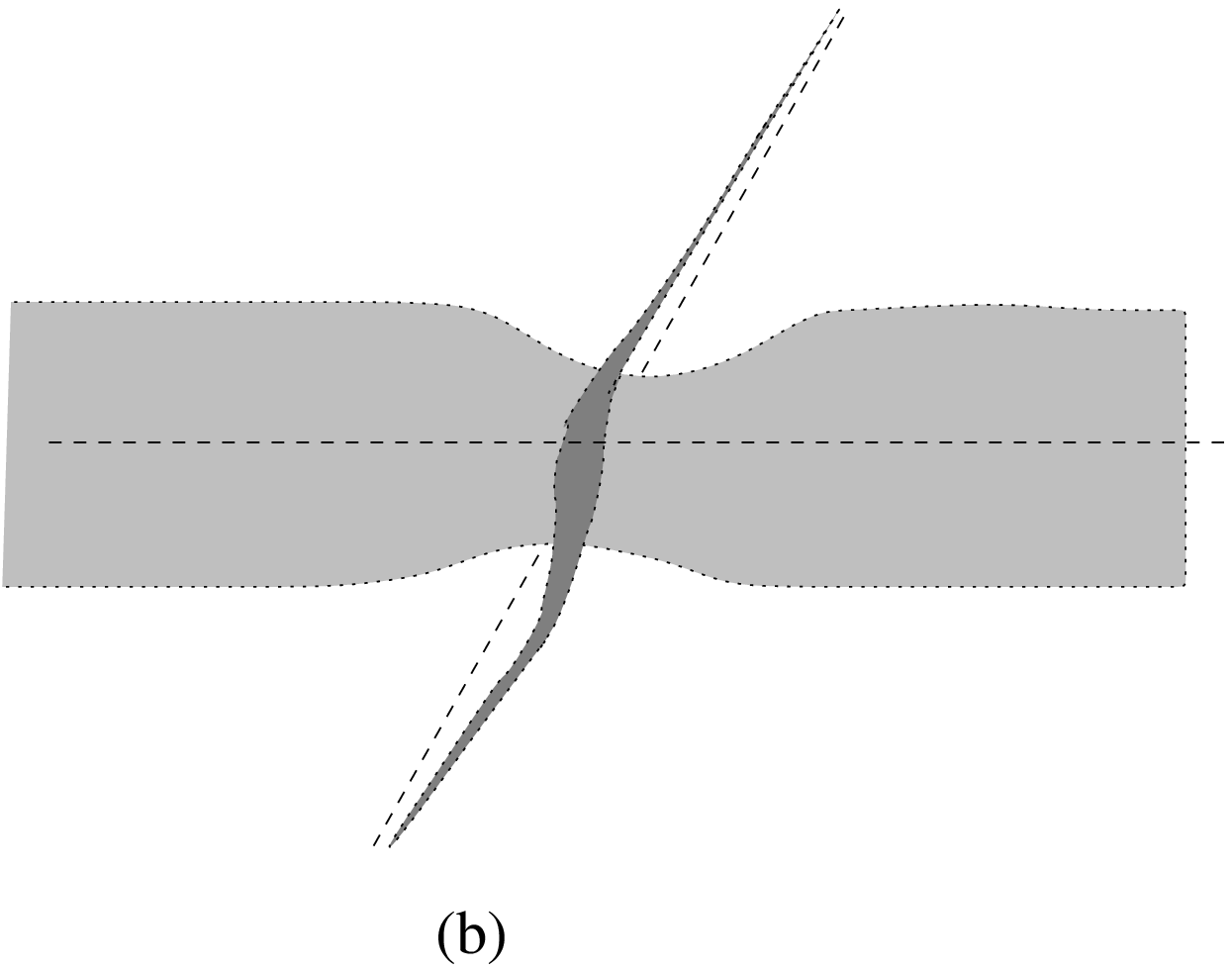}}{\at{5.2cm}{1.7cm}{$\delta$}\at{1.4cm}{3.3cm}{\boxit{\hbox{$\
          \kappa<\kappa_c\
      $}}}\at{0.7cm}{1.3cm}{$\zeta_{\rm phot}$}\at{2cm}{0.7cm}{$\zeta_{\rm atom}$}}}
$$\hbox{\box71\box72 \raise 0.1cm \box73}$$
Figure 7. Variation of the resonances' energies with respect to the detuning, at the  

\hskip 0.85cm transition between the strong
  (a) and weak (b) coupling regimes.

\hskip 0.85cm A real representation.
\medskip


\noindent
shows (see below) that for a certain $\kappa_c$, the two line segments at $\delta=0$ coincide (same energy at the center and same length, i.e. the resonances coincide in the complex plane). When $\kappa$ decreases and crosses this value, figure 7a changes in a continuous way into figure 7b.

The two regimes called "strong coupling regime" and "weak coupling regime" in the literature clearly appear in this picture, on each side of $\kappa_c$. In the 
weak coupling regime, the labelling $\zeta_{\pm}$ is no longer pertinent. However, one of the two resonances can be associated with the atom and the other one with the photon. This is translated in the notations $\zeta_{\rm phot}$ and $\zeta_{\rm atom}$. When the coupling increases beyond the critical value, this labelling is no longer possible. The mixing of the states is important when the detuning is close to zero. An atomic state is changed continuously into a photonic state when the detuning increases from $-1$ to $1$.

The preceding description gives a picture of the displacement of the two resonances in the complex plane. For $\mu=0.01$, the critical value of $\kappa$ is close to $3\times 10^{-3}$.\break 

\vskip 0.8cm
\newbox\billion
\setbox\billion=\hbox{\boxit{\vtop{\vskip0.2cm  \hbox{$\
        \kappa=0.0031\ $}\vskip0.2cm}   }  }

\newbox\billionnet
\setbox\billionnet=\hbox{\boxit{\vtop{\vskip0.2cm\hbox{$\ \kappa=0.0029\ $}\vskip0.2cm}}}

\setbox81=\hbox{\psannotate{\psboxto(6cm;0cm){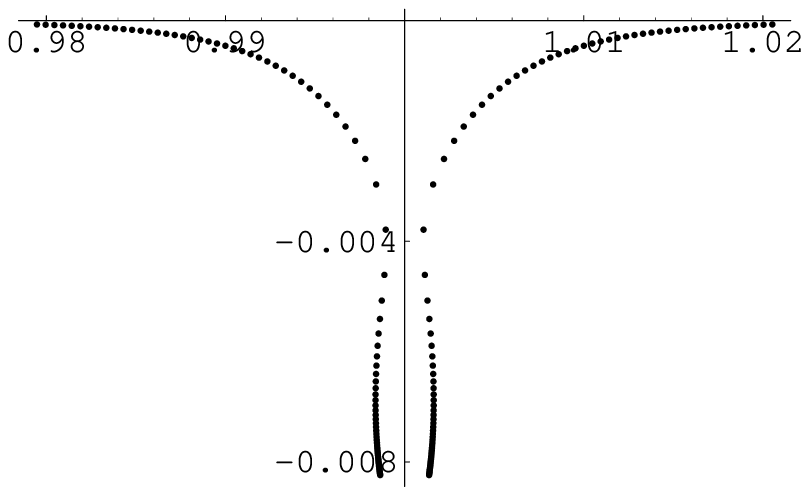}}{\at{2.2cm}{4cm}{\box\billion}\at{1cm}{-0.3cm}{$\delta=0.02\nearrow$}\at{3.4cm}{-0.3cm}{$\nwarrow\delta=-0.02$}\at{1.5cm}{2cm}{$\delta=0\rightarrow$}\at{3.4cm}{2.1cm}{$\leftarrow\delta=0$}\at{-0.7cm}{3cm}{$\delta=-0.02$}\at{5cm}{3cm}{$\delta=0.02$}\at{3.5cm}{1.3cm}{$\zeta_+$}\at{2.3cm}{1.3cm}{$\zeta_-$}}}

\setbox82= \vbox{   \boxit {  \vbox{\vskip0.2cm\hbox{ $\mu=0.01\
        $}\vskip0.2cm}}    \vskip 1.5cm}

\setbox83=\hbox{\psannotate{\psboxto(6cm;0cm){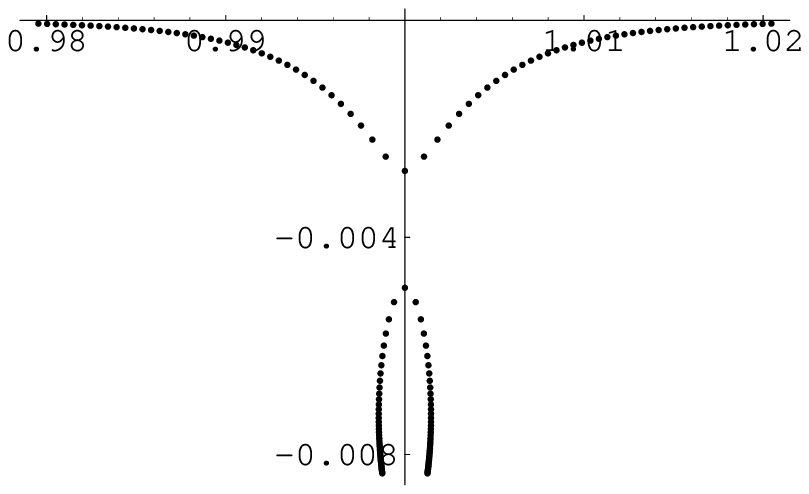}}{\at{2.2cm}{4cm}{\box\billionnet}\at{1cm}{-0.3cm}{$\delta=0.02\nearrow$}\at{3.4cm}{-0.3cm}{$\nwarrow\delta=-0.02$}\at{1.5cm}{1.4cm}{$\delta=0\rightarrow$}\at{3.4cm}{2.3cm}{$\leftarrow\delta=0$}\at{-0.2cm}{3.8cm}{$\delta=-0.02$}\at{5cm}{3cm}{$\delta=0.02$}\at{4cm}{3cm}{$\zeta_{\rm
        atom}$}\at{3.4cm}{1.1cm}{$\zeta_{\rm phot}$}}}

$$\hbox{\box81\hskip -0.5cm\box82\hskip -0.5cm\box83}$$
\smallskip
Figure 8.  Variation of the resonances' energies with respect to the detuning, at the

\hskip 1.6cm  transition between the strong and weak coupling regimes.

\hskip 1.6cm Representation in the complex plane of $\zeta=E/(\hbar ck_0)$.

\bigskip
In figure 8, we show the exact position of each resonance, as a function of the detuning, in the two regimes. The dotted parts of the curves are obtained through varying $\delta$ step by step; the points have not been joined by a curve so as to underline the rapid variation near the singular point.

From figure 8, it is clear that, for $\delta=0$, $\zeta_{\rm atom}$
is standard whereas $\zeta_{\rm phot}$ is not, since the curve $\zeta_{\rm atom}$ is going to flatten on the reals as $\kappa$ decreases. Let us also note that the only knowledge of the resonances at $\delta=0$ does not allow the regime to be determined. (Incidentally, their real parts are not exactly $1$.) In the weak coupling regime, the curves representing the resonances' real parts cross as the detuning varies, whereas we have an anti-crossing in the strong coupling regime. The distinction between the two regimes requires a detailed analysis near $\delta=0$.

One could also look at the regime's transition through varying $\mu$, $\kappa$ being fixed. The strong coupling regime would then occur below a critical $\mu$, depending on $\kappa$. Note that the ratio $\kappa_c/\mu$ is about $0.3$ for the $\mu$ value we considered.

We take this example as a model in defining the two regimes for the coupling of a two level system ${\cal S}$ to a continuum.

\underbar{\sl Definition {\rm 2}}: {\it We say that we are in a strong coupling regime if, through the variation of the detuning between ${\cal S}$ and the continuum, the two resonances move as it is represented in figure {\rm 7a} or figure {\rm 5}. We are in a weak coupling regime if the two resonances move as it is represented in figure {\rm 7b}.}

Note that this definition depends on the existence of a parameter measuring the detuning and that it therefore does not apply when this parameter is no longer obvious.
In section 4, we introduce the coupling function  $-\sqrt 3 \ \big(1+(k/k_0)^2\big)^{-2}\ k/k_0$, the atom level-spacing being $1$; in this case there is no obvious detuning parameter.

When the continuum is coarsely discretized into three levels as before, there is no critical $\kappa$, but the atomic states gradually shows up when the coupling constant decreases. For instance, figure 9 describes what figure 6 becomes when $\kappa=0.002$.
\medskip

$$
\psannotate{\psboxto(6cm;0cm){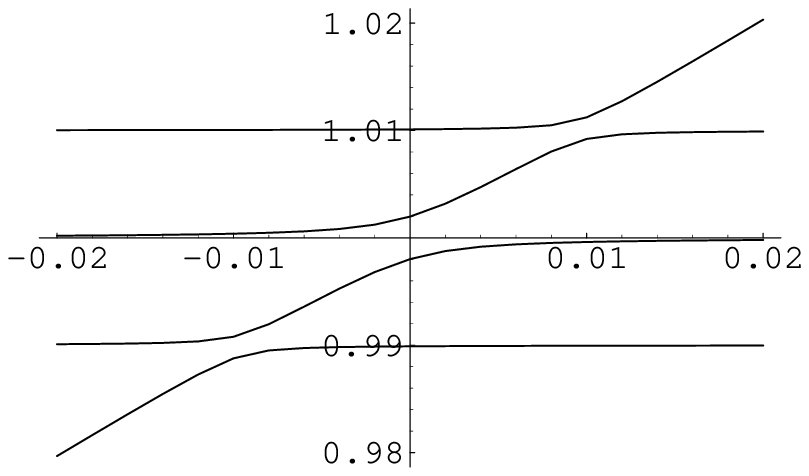}}{\at{6.2cm}{1.8cm}{$\delta$}\at{2.45cm}{3.9cm}{$E/\hbar ck_0$}}
$$
\centerline{Figure 9. Variation of four eigenvalues, at $\kappa=0.002$, for a discretization of the} 

\hskip 1.7 cm continuum. $\mu=0.01$.


\medskip
2.4.3.2. {\it The $\lambda\rightarrow 0$ behaviour of resonances sitting at $\zeta_{w,ph}(\mu)$ and $\zeta_{w,at}(\mu)$. Standard and nonstandard resonances.} It can be checked by computer, up to $\mu=5$, that $\zeta_{w,ph}(\mu)$
is nonstandard and that $\zeta_{w,at}(\mu)$ is standard.

\smallskip
2.4.3.3. {\it Behaviour of resonances sitting at $\zeta_{w,ph}(\mu)$ and $\zeta_{w,at}(\mu)$
  when the coupling increases.} Let us consider the resonance which sits at $\zeta_{w,ph}(0.01)$ when $\kappa=0.1$ and follow its trajectory as $\kappa$ increases. We find that its real part decreases and that its imaginary part also decreases, down to a value around $0.11-2\times  10^{-6}\ i$ for $\kappa\simeq 1.0062$. Let us call ${\cal R}$ the curve segment thus drawn. This reminds us of the behaviour that we mentioned in the introduction, a behaviour that we illustrated elsewhere (Billionnet 2002), with the function $g(p)=\sqrt{(2/\pi}\ p/(1+p^2)$: in that case, the analogous resonance became real negative beyond a certain value of $\kappa$. The existence of this eigenvalue has been known for a long time. In the present case the resonance does not become real; its imaginary part starts growing beyond the above-mentioned value of $\kappa$. Nevertheless, for large $\kappa$, greater than 1.2 for instance, one can see that there does exist a negative eigenvalue of the Hamiltonian. It approaches $0$ when $\kappa$ decreases to $1.118$ and connects to ${\cal R}$, but only if $\kappa$ follows a path avoiding a neighbourhood of $1.1$, in the complex plane. This indicates a branch point of the zeros of the multivalued function $f$, in this region of the $\kappa$ complex plane. A complication may be due to the following fact. There is a difficulty for a zero of $f$ to cross $0$: for $\zeta\sim 0^-$, the integral in (8') diverges, because $g$ does not vanish at $0$, contrary to the above-mentioned case. It would be interesting to multiply the coupling function (7) by $p$ and see whether the negative eigenvalue reaches $0$, which is likely to be the case.

\noindent
The resonance which sits at $\zeta_{w,at}(0.01)$ for $\kappa=0.1$ tends to the positive real axis at infinity, after having moved away from it for a while.

We are now going to go through several subjects concerning mixed states (also called intricate, dressed or hybrid states in the literature) and the two coupling regimes, for interactions of a discrete-level system with a continuum. These subjects have been often studied in the last years. In each of these examples, the three parameters we used before will come into play.

\bigskip
{\bf 3. Illustration of the general study in several concrete cases.}  

\smallskip
{\bf 3.1. Atoms (or equivalents) in cavities. The continuum is a continuum of photon states}
\smallskip

Let us consider an atom with two levels (states $\ket 0$ and $\ket 1$), in resonance with the mode of a cavity with quality factor $Q$. Let $\omega$ be the angular frequency of the photons. We refer to the introduction for some works on cavity electrodynamics. One may also consult S. Haroche's courses at College de France (2001-2004), which are accessible, for recent studies. The strength of the atom-mode coupling is measured by the frequency $\Omega={1\over \hbar}\ D_{10}\sqrt{\hbar\omega/(2\epsilon_0V)}$, $D_{10}$ being the matrix element of the electric dipole operator between the two states and $V$ the cavity's volume. Since the cavity is not perfect, the mode may be described as an environment presenting a Lorentzian spectrum with width $\Gamma_c=\omega/Q$ to the atom. When $Q$ is large, we are in the small $\mu$ case of the general presentation and thus in a strong coupling regime. In the $\Gamma_c=0$ limit, and with at most one photon present, we are in the case of figure 3: the coupling yields intricate atom-photon eigenstates (polariton states), which are linear combinations of  $\ket{0,{\rm 1\ photon}}$ and $\ket{1,{\rm 0\ photon}}$ states. If $\Gamma_c$ is not considered as zero, these states become the two resonances described in the general presentation. We are in the case of figure 5. The width at infinity on the abscissa axis is $2\Gamma$. The complex values of the resonances, and in particular their imaginary parts, vary with the detuning. When the detuning changes from a large negative value to a large positive one, a ``photonic state'' changes continuously into an ``electronic state'' or conversely, depending on which resonance is considered. We get there intricate states which have been much studied these last years. Rydberg atoms are specially appropriate to the study of this atom-cavity strong coupling (Haroche 1984). Studies are presently conducted on this subject. It might be useful to look at the resonances' position in the complex plane to get a more precise description than the approximation given by figure 3. 

In section 4 we examine what becomes of the resonances in a case where the cavity is no longer present.

We find an analogous situation for excitons in semiconducting microcavities (Weisbuch 1992, Yamamoto {\it et al} 2000, Senellart 2003). One can vary the photon continuum's width (through changing the cavity's quality factor) or the coupling strength (for example by means of a magnetic field (Yamamoto {\it et al} 2000 p. 43)), so as to pass from the strong to the weak coupling regime. In the former, the experimental curves show two peaks, whereas in the latter, they often show only one peak. Since we did find two resonances in both regimes, we must explain this. Three reasons may be put forward. The first one is that one resonance's imaginary part may become large (widening of the peak which disappears). The second is that one resonance may stay out of the energy range which the apparatus tests. A third possibility is that the probe be sensitive to the cavity's state and not to the exciton's state (see (Haroche 1992 Sect. 3.3), for the atom-and-cavity case).

The excitons may also acquire a certain width, through exciton-exciton or exciton-phonon interactions, but we do not consider interactions of two continuums in this paper.

\medskip
{\bf 3.2. Excitons in a microcavity. A case where the continuum consists of excitonic states}
\smallskip

Let us now assume that the photon in the cavity is practically monochromatic. We may have the exciton's energy spectrum seen by the photon varying by means of a magnetic field. The spectrum has a discrete part and a continuous part and both changes with the magnetic field. When the photon energy is in the discrete part of the exciton's spectrum, we get several possible resonances, with small width, and the anti-crossings when two different exciton-photon states have neighbouring energies. When the photon energy is in the continuous part of the exciton spectrum, the photon energy gets widened (Tignon 1995).

The first situation corresponds to the strong coupling regime of figure 3, for each anti-crossing. Both resonances at stake at each anti-crossing have a small width. None of them can be associated with the photon or with the exciton if the detuning is not large. Corresponding mixed states are called magnetopolaritons. In the second situation, expe\-rimental curves show that the coupling to the continuum  gives the photon state a certain width (Tignon 1995 fig.2). Through calculating the displacement in the complex plane of the photonic resonance, as the detuning passes from a negative value (not too big, the photon's energy has to stay in the continuum) to a positive one, one should get the same picture as for $\zeta_{\rm atom}$ in figure 7b. 

\medskip
{\bf 3.3. Electron-phonon mixed states}

\smallskip
An other example where ideas developed in section 2 apply, although with some modifications, is the coupling of electrons confined in quantum dots to longitudinal optical phonons of a bi-atomic lattice. The interaction is that of the electron with the electric field created by the lattice dipoles, a field which oscillates according to the various possible modes. The Fr\"olich Hamiltonian of the electron-phonon system formally reads
$$
H_{\rm el-ph}=\sum_{n\geq 0} E_n \dyad n n\otimes 1+1\otimes \sum_{{\bf k}\in {\cal B}}
\hbar\omega(k)\ a^*_{\bf k} a_{\bf k}+ {\cal N}^{-{1/2}}C\ \sum_{{\bf k}\in {\cal B}}
k^{-1}\ (a_{\bf k}^* e^{-i{\bf k}
\cdot {\bf r}}-a_{\bf k}e^{i{\bf k}\cdot{\bf r}})              \eqno (11)
$$
where $\ket n,\ n=0,1,\cdots$ denotes the eigenstates of the electron in the dot,  $\omega(\bk)$ is the energy of a phonon with wave number $\bk$, ${\cal B}$ is the first Brillouin zone, $ C$ is a pure imaginary constant and ${\cal N}$ a normalization factor (Callaway 1974 p. 656). If the lattice is infinite, the possible values of ${\bf k}$ make a continuum and the Hamiltonian may be written
$$
H_{\rm el-ph}=\sum_{n\geq 0} E_n \dyad n n\otimes 1+1\otimes H_{\rm phon}+
\sum_{m,n}\lambda_{mn}\big(\dyad m n\otimes a^*(\overline g_{mn})+\dyad n m\otimes a(g_{mn})\big)                                                       \eqno (12)
$$
where
$H_{\rm phon}$ is the energy operator in the phonon space:
$H_{\rm phon}=\hbar\int_{\cal B}\omega(k)a^*_{\bf k}a_{\bf k}\ d{\bf k}$ and 
$$
g_{mn}({\bf k}):=i\ \Big(\int_{\cal B}|\obraket  m{k^{-1}\ e^{i{\bf k}\cdot {\bf
        r}}}n|^2\ d{\bf k}\Big)^{-1/2}\times \obraket m{k^{-1}\ e^{i{\bf k}\cdot {\bf r}}}n                                             \eqno (13)
$$
are the (normalized) coupling functions; the coupling constants
$$
\lambda_{mn}=(2\pi)^{-3/2}\ i\ C\ \Big(\int_{\cal B}|\obraket  m{k^{-1}\ e^{i{\bf k}\cdot {\bf
        r}}}n|^2\ d{\bf k}\Big)^{1/2}                  \eqno (14)
$$
have the dimension of an energy. 
In calculating eigenvalues of $H_{\rm el-ph}$, one often limits oneself to considering two particular levels $\ket 0$ and $\ket 1$, for
example the first two levels of the dot, also neglecting the Hamiltonian's matrix elements which are of order greater than one. In
(Hameau 1999), one of the strong coupling regime exhibited involves states (s, 1 LO phonon) and (p, 0 LO phonon) for electrons in InAs quantum dots. Let us consider this approximation and this example. Function $\omega({\bf k})$ is maximum for ${\bf k}=0$, where
it is equal to 36 meV. The range of $\omega({\bf k})$ is 8 meV. Actually, if we take account of the $k^{-1}$ dependence of the interaction and limit ourselves to ${\bf k}$'s which give appreciable values of the matrix elements, the range reduces to 0.4 meV. The electron-phonon detuning is obtained through varying the quantum levels in the dot by means of a magnetic field. Let us denote the level spacing by $E(\delta):=\hbar\omega_0(1+\delta)$, 
$\delta$ measuring the detuning with respect to the photon energy mean value $\hbar\omega_0$. The Hamiltonian is
$$
H=E(\delta)\dyad 11\otimes 1+1\otimes H_{\rm phon}+\lambda \big(\dyad 10\otimes\  a^*(\overline g)+\dyad 01\otimes\  a(g)\big)                \eqno (15)
$$
where $g =g_{01}$ and $\lambda=\lambda_{01}$.
If $\omega(\bk)$ had only one value $\omega_0$, the eigenvalues would be roots of equation $z-E(\delta)-\lambda^2/ (z-\hbar\omega_0)=0$. Indeed, eigenvalues or resonances are obtained by means of the function
$$
f(\lambda,\delta,z):=z-E(\delta)-\lambda^2\int_{\cal B}{ |g({\bf
    k})|^2\over z-\hbar\omega({\bf k}) }\ d{\bf k}  \ .    \eqno (16)
$$
They are its zeros or those of some analytic continuation in the lower half-plane. When the phonon continuum is infinitely narrow, the zeros are therefore real and there variation with $\delta$ is of the type shown in figure 3. When the width is no longer $0$, it is interesting to see whether resonances are described by figure 5 or figure 7b, i.e. whether the coupling regime is strong or weak. Interpreted in the subspace spanned by states (s, 1 LO phonon) and (p, 0 LO phonon), data given in (Hameau 1999) for the energies of stationary states, or almost stationary states (see the remark just below and in the next to last paragraph of this section), show that we are not in the case of figure 7b, but in a strong coupling regime. As a consequence, when $\delta$ passes from a large negative value to a large positive one, a phonon state changes continuously into an electronic state. Let us note that the smallness of the continuum's effective width, and also the limit in the measurements' precision imply that points which should be represented in a figure of the figure 5-type are actually represented as in figure 3.

There is a difference between this problem and that of the coupling of a discrete system to the photon. In the present case there are two functions which contribute to the resonances' imaginary part: $g$ and $\omega$. In the limit $\delta\rightarrow\infty$, the vertical extension of the lower surface of figure 5, which expresses the imaginary part of one of the resonances, depends on both widths. Only explicit calculations would tell us how the widths of these two functions (and even the functions themselves) contribute to the final result.

Unfortunately, a numerical calculation is more complicated than in the case where $\omega(\bk)=|k|$. It has not been done. Indeed, performing an analytic continuation  requires knowing the values $\bk_1(z),\bk_2(z),...$ for which $\omega(\bk)=z$. Even in the case where $\omega(\bk)$ has an explicit form, the $\bk_i(z)$ are not simple functions. For example, for a one-dimension lattice with equal mass atoms, the $\bk_i(z)$ are of the form $arcos(c\ z)$, a function which is multivalued. Nevertheless, let us show qualitatively how $\omega(.)$ may create a resonance distinct from the one which is close to $E(\delta)$ when the coupling constant is small. In one dimension, and in the case of equal mass atoms, we have $\omega(k)=\omega_{\rm max}\  \cos(a k/4)$, with $a$ the lattice spacing. Therefore, values taken by $\omega$ lie in $I=]{\sqrt 2\over 2}\omega_{\rm
  max},\omega_{\rm max}]$. When $z$, coming from the upper half-plane, crosses this interval at a point different from $\hbar\omega_{\rm max}$, the integrand's denominator in (16) has two poles in the integration interval, $]-\pi/a,\pi/a]$, which are $\pm arccos((\hbar\omega_{\rm max})^{-1}\ z)$; the integration interval can be deformed so as to avoid these two poles whereas this is not the case if $z$ comes to $\hbar\omega_{\rm max}$, the integral becoming divergent. $z=\hbar\omega_{\rm max}$ is thus a singularity. Let us assume it be the only one and, moreover, a simple pole. By analogy with the expression $z-E(\delta)- C^{\rm te}\
  \lambda^2/(z-\hbar \omega_{\rm max})$, we may expect a zero of $f(\lambda,\delta,.)$ near $\hbar \omega_{\rm max}$, for $\lambda$ small. The zero of $f(\lambda,\delta,.)$ will in fact be complex because the continuation of that function is complex. 
Let us recall other singularities of the continuations of $f(\lambda,\delta,.)$, already met in the photonic case. They are due to poles of $g$. For example, if $k_p$ is a pole of $g$ in the lower half-plane, $\omega(k_p)$ may be a singular value of some analytic continuation of $f(\lambda,\delta,.)$.

Nevertheless, from general ideas deduced from the analysis of section 2, one can make two points. Firstly it is because the continuum is narrow that the strong coupling regime occurs, the coupling then resembling that of discrete states, with real energies. Secondly, there is an important difference with this latter case: the width of the phonon states' continuum, as small as it may be, makes the mixed electron-phonon states unstable, since the energies of these states now have a nonzero imaginary part. In the same way, the photons, although they are stable, give an imaginary part to the electron energies of the naked atom, through the extension of their spectrum. This remark may be useful in discussing the stability of polarons.

An analogous situation occurs in the case of the exciton-phonon coupling in semi-conductor quantum dots (excitonic polarons)(Verzelen {\it et al} 2000, Verzelen {\it et al} 2002).

\vfill\eject
{\bf 4. Standard and nonstandard resonances involving large-$n$ states of the hydrogen atom}
\smallskip
We are now going to take more specifically into account the fact that the environment seen by a system ${\cal S}$ depends on the state in which the system is. The hydrogen atom is a first example of a system about which one can answer the following general questions. When one considers atomic or molecular transitions between states whose spatial extension increases, does one see any decrease in the imaginary part of some of the associated resonances? In what conditions would the order of magnitude of the imaginary parts of the standard and nonstandard resonances be comparable? To ask these questions is justified by the example of the charged harmonic oscillator studied in (Billionnet 2001). Let us recall the result. If physical parameters of the oscillator have such values that the spatial extension of the wave functions is large enough compared to the wavelength of the fundamental transition, then the nonstandard resonance may become a (real) negative eigenvalue, therefore corresponding to a stable state. We want to set a calculus for extended states of the hydrogen atom, with a parameter measuring the ratio between the space extension and the transition's wavelength and calculate an example of a nonstandard resonance. This study will also give us an opportunity to give a new example of a coupling function, in a case where no exterior constraint is applied on the atom-field system. In the cavity case, this constraint existed; it could suppress or enhance an atomic transition.

\medskip
{\bf 4.1 Setting of the calculus and introduction of non-dimensional variables}

Let us consider a transition between two states $\ket 1=\ket{n_1,l_1,m_1}$ and $\ket
2=\ket{n_2,l_2,m_2}$ of the electron in the atom, accompanied by the emission of a photon. The space of possible photon states is assumed to be the space generated by states $\ket\gamma=\ket{k,j,m,\lambda}$, with variable energy $E=\hbar ck$, the angular momenta $j,m$ and the polarization $\lambda$ being fixed. The normalization is  $\braket{E,j,m,\lambda}{E',j',m',\lambda'}=E\delta(E-E')\delta_{j,j'}\delta_{m,m'}\delta_{\lambda,\lambda'}$. Taking $H_I=i\ e\hbar/(mc)\  {\bf A}.{\bf \nabla}$ as the interaction Hamiltonian and assuming $j+j_1+j_2$ to be for example even, we have (see for example (Moses 1973)) 
$$
g_I(k):=\obraket{1}{H_I}{2,\gamma}=C\big( A\ \phi_1(k)+ B\ \phi_2(k)\big)  \eqno (17)
$$
where
$$
\phi_1(k)=\int j_j(kr) R^*_1(r)\big({d\over dr}R_2(r)\big)\ rdr\ ,\quad \phi_2(k)=\int
j_j(kr)\big({d\over dr} R^*_1(r)\big)R_2(r)\ rdr  \ .   \eqno (18)
$$
Constants $A$ and $B$ depend on the two considered states $\ket 1$ and $\ket 2$ and $R_1$, $R_2$ are the radial parts of their wave function.  Let us neglect matrix elements  of $H$ which are not in the subspace generated by $\ket 1$ and $\ket{2,\gamma}$. Let $H'$ be the corresponding operator, acting in this subspace. The distance, which we call $z$, between eigenvalues or resonances of $H'$ and the fundamental energy is one of the zeros of the following function
$$
f(z)=z-{\cal E}_{n_1,n_2}-2\ ||g_{I}||^{2}\ \int_0^\infty{|g(k)|^2\over z-\hbar ck}\
{dk\over k}                  \eqno        (19)
$$
or of its analytic continuation into the lower half-plane.  $g(k)$ is $||g_{I}(k)||^{-1}\  g_{I}(k)$, with
$||\theta||=\Big(2\int_0^\infty |\theta(k)|^2\ dk/ k\Big)^{1/2}$ and ${\cal E}_{n_1,n_2}$ is the difference between the energies of states $\ket 1$ and $\ket 2$. The $k^{-1}$ factor comes from the normalization of $\ket{E,j,m,\lambda}$. In preceding works, we studied the zeros of multivalued functions of the same form but with other $g$'s. Before we give indications on the form that $g$ takes here in some particular transitions, let us show that the equivalent of parameter $\mu$ of section 2.4 is now the ratio of the atomic transition wave length to a length measuring the space extension of states  $\ket 1$ and $\ket 2$.
We have 
$$
R_1(r)=P_{n_1,l_1}(r/ a_0)\ \exp(-r/( n_1a_0))\ ,\quad
R_2(r)=P_{n_2,l_2}(r/ a_0)\ \exp(-r/( n_2a_0))
$$
where $P_{n_i,l_i}$ are polynomials and $a_0$ is the Bohr radius.
Let us introduce $\overline \rho_{n_1,n_2}=\big((n_1a_0)^{-1}+n_2a_0)^{-1}\big)^{-1}$, half the harmonic mean of the extensions $n_1a_0$ and $n_2 a_0$ of $\ket 1$ and $\ket 2$. 
Let us set $y=2\pi\overline \rho_{n_1,n_2}/\lambda_{\rm phot}$ and
 $G(y)=g(y/\overline\rho_{n_1,n_2})$. We have $||G||=1$.
Through also introducing the non dimensional variable
$\zeta=z/{\cal E}_{n_1,n_2}$, $f(z)$ changes into ${\cal
  E}_{n_1,n_2}\ F(\zeta)$, with
$$
F(\zeta):=\zeta-1-2 \kappa^2\int_0^\infty{|G(y)|^2\over
    \zeta-\mu y}\ {dy\over y} =  \zeta-1-2 \kappa^2\int_0^\infty{|G_\mu(y)|^2\over
    \zeta- y}\ {dy\over y}             \eqno (20)
$$
where 
$$
\kappa={\cal E}_{n_1,n_2}^{-1}\ ||g_{I}|| \quad {\rm and}\quad \mu=\lambda_{n_1,n_2}/(2\pi\ \overline\rho_{n_1,n_2})         \eqno (20')
$$
$G_\mu(y)=G(\mu^{-1}y)$ being obtained from $G$ through the unitary dilation operator in\break $L^2(\bbbr,dy/y)$. 

\medskip
{\bf 4.2 Comparison between the standard and the nonstandard resonance. Dependence with respect to the spatial extension of the naked states.}

The study of the standard and nonstandard resonances has been changed into the study of the zeros of the multivalued function $F$. Through comparing (20) with (8) we see that parameter $\mu$ here plays the same role as in section 2.4: it dilates the coupling function. However, it is not exactly the same dilation.

If $E_I$ is the atom's ionization energy, we have 
$$
\mu=(n_1-n_2)^{-1}\ n_1n_2\
E_I^{-1}\ {\hbar c/a_0}>E_I^{-1}\ {\hbar c/a_0}\simeq 2/\alpha\simeq 274\ .
$$
We are going to show that the two zeros giving the standard and the nonstandard re\-sonances, are respectively close to $1$ and $-i\mu$, if $\kappa$ is small, which we will check in some examples in section 4.3. We then get a qualitative answer to the two questions we asked at the beginning of section 4. Firstly one of the resonance does move towards the reals when the mean extension of the states increases. Secondly, and this is eventually
the important point in the present case, $\mu$ is large and, therefore, the nonstandard resonance sits much farther from the real axis than the standard one and can be ignored. That there is a zero near $1$ is clear. Let us show that there is a zero near $-i\mu$.

Function $G$ depends on the states $\ket 1$ and $\ket 2$ but its poles do not depend on them. Let us show this if $j=1$. Introducing  $x:=\overline\rho^{-1}\ r$, we get $\Phi_i(y):=\phi_i(y/\overline\rho)=\int_0^\infty j_1(yx)P_i(x)e^{-x}\ dx$, where $P_i$ is a polynomial whose degree is at least $2$ and at most $n_1+n_2-1$. Setting $A_p(y):=\int_0^\infty j_1(yx)x^p e^{-x}dx=(-1)^{p+1}\big(x^{p+1}{d^p\over dx^p}(x{\rm Arctg}(1/x))\big)_{x=1/y}$, we get, for $p\geq 2$,
$$
A_p(y)=(-1)^{p+1}{y\  Q_{p-2}(y)\over(1+y^2)^p}    \eqno (21)
$$
where $ Q_{p-2}$ is a polynomial with degree at most $p-2$. As a consequence, $G(y)$ has the form $y\sum_{p\geq 2} a_p(1+y^2)^{-p}\ Q_{p-2}(y)$. Therefore, it has two poles and only one in the lower half-plane, at $y=-i$. 
This implies that the analytic continuation of $F$ into the lower half-plane has a pole at $\zeta=-\mu\ i$, since the integration contour in (20) is pinched between $\zeta/\mu$ and $-i$, poles of the integrand. It is this pole of $F$ which is important for the nonstandard resonance. (This was already the case in section 2.4, since the position of the pole of $f$ in (8') was related to the width of $g$ through the position of the pole of $g$.) Indeed, for small $\kappa$, the analytic continuation $F_+$ of $F$ into the lower half-plane has a zero, say $\zeta_{\rm n.s.}$, near this pole $-\mu i$. (Think of the function $\zeta-1-\kappa^2 (\zeta-a)^{-1}$ which has a zero near $a$.)

Of course the exact position of the zero of $F$ associated with the nonstandard resonance does not depend only on the pole of $G$ but also on the exact form of $G$. In particular the position also depends on the order of the pole: clearly, as the order increases, the zero gets farther from the pole. In the numerical example we give in section 4.3 below, the order of the pole is two. It is larger for other transitions.
But on some examples we saw that increasing the order does not seem sufficient to move the nonstandard resonance substantially closer to the real axis. 

In conclusion, large-$n$ states of the hydrogen atom do not give any other interesting resonances than the standard ones. Let us compare this result to the one we obtained for the extended system mentioned at the beginning of section 4. We considered a quantum charged harmonic oscillator with charge $1$, mass $m$ and spring constant $k_r$. The level spacing is $\hbar \sqrt{k_r/m}$ and the exponential decrease of the wave functions is $\exp(-r^2/\delta^2)$, with $\delta=\hbar^{1/2}(k_rm)^{-1/4}$. The larger $\delta$, the larger the extension. In a model in which this oscillator is coupled to the transverse electromagnetic field, we saw (Billionnet 2001) that the nonstandard resonance moves towards the reals when $2\pi\delta/\lambda=c^{-1}\hbar^{1/2}(k_r/ m^3)^{1/4}$, the equivalent of $1/\mu$, increases. In particular, this resonance becomes even real negative if the ratio $2\pi\delta/\lambda$ becomes larger than $3\sqrt{2\pi}/\alpha$. In the hydrogen atom case, the ratio $2\pi\overline\rho/\lambda=\mu^{-1}$ remains smaller than $1/274$, and this implies that the nonstandard resonance always remains far from the reals. As regards the second question asked in the beginning of the section, it has to be noted that the distance of the standard resonance to the real axis is proportional to $\kappa^2$. Therefore, in the present study, since $\kappa$ is small (see below), $\mu$ would have had to be much smaller than $1$ in order that the nonstandard and standard resonances had comparable imaginary parts.

\medskip
{\bf 4.3 A numerical example}

We might have calculated the nonstandard resonances, in the two-level approximation, for transitions $n_1\rightarrow n_2=1$, for which $\mu$ is close to its lower bound 274. However, in order to give an idea of the position of such resonances, it is sufficient to perform the calculation in the simpler case of the transition $\ket 1=\ket{2,1,0}\rightarrow\ket 2=\ket{1,0,0}$, for which $\mu$ is only twice the lower bound. The calculus is given in Appendix B. Here is the result.

The zeros of $F$ are respectively $\zeta_{2,{\rm s}}\simeq 1-2\times 10^{-6}-2\times 10^{-8}i$ and $\zeta_{2,{\rm n.s.}}=1.493-544\ i$. The former gives the standard resonance: $z_{2,{\rm s}}={\cal E}_{2,1}\ \zeta_{2,{\rm s}}$. Its imaginary part gives a life time $\tau_2=2\times 0.16\times 10^{-8}$ s, which has to be divided by $2$ to take the other polarization into account. We thus recover the life time of the $2p$ state. The other zero is the nonstandard resonance, very far from the real axis.

We have also calculated $\kappa$ for the transitions between $\ket 1=\ket{n,n-1,0}$ and $\ket 2=\ket{n-1,n-2,0}$, with a photon in a state $(j,0,+1)$. The result, given in Appendix B, shows that $\kappa$ remains of the order of $0.02$, when $n$ varies between $10$ and $50$. Thus the non-dimensional coupling constant does not increase although ${\cal E}_{n_1,n_2}^{-1}\simeq n^3\ E_I^{-1}$ gets large in (20').

In order to get the exact position of the resonances, one should of course take other transitions into account (see (Billionnet 2005)). But we do not see any reason why this should substantially displace the nonstandard resonances towards the real axis .

Regarding extended systems, we could think of hydrogenic excitons, whose mean radii may be as large as $2000\  \build {\rm A}_{}^{\circ}$. But the associated Rydberg constant and Bohr radius are respectively $Ry^*=\epsilon_r^{-2}\ (m_{\rm red}/ m_0)\ Ry$ and $a_B^*=\epsilon_r\ (m_0/ m_{\rm red})\ a_B$, $m_{\rm red}$ being the reduced mass of the electron-hole system, $m_0$ the electron's mass in the vacuum and $\epsilon_r$ the medium's relative permittivity (Weisbuch and Vinter 1991, formula (20a)); therefore, the ratio $\mu$ remains large. 

\bigskip
{\bf 5. Conclusion}

\smallskip
The analysis in the complex energy plane of the resonances of a system ${\cal S}$ coupled to a continuum gives a precise mathematical description of mixed states which forms in the interaction. These mixed states may be assimilated to eigenstates of the Hamiltonian if the continuum is very narrow but, in general, they have a nonzero imaginary part, which may be large. When their imaginary part is so small that it can be considered to be zero, the corresponding eigenstates mix states of ${\cal S}$ with states of the continuum. It is these mixed states which are important in certain situations, for example in some spectroscopic measurements. When the imaginary part is not negligible, one has to deal with resonances, which we will still consider as associated to mixed (unstable) states. The description we get is more complete than the perturbative one based on the unperturbed states of ${\cal S}$.

These mixed states may be followed with respect to various parameters describing  ${\cal S}$ or the coupling. When the parameter is the detuning, the description allows us to give a precise definition of the strong and weak coupling regimes. Data yield numerous examples of these mixed states when the continuum is narrow. Mixed states in the sense of the above paragraph also exist for the hydrogen atom coupled to the transverse electromagnetic field; but we saw that only the usual ones, corresponding to the unstable atomic levels, have a small imaginary part. They correspond to the resonances we called standard. A condition for nonstandard resonances to play a role for a system like the preceding atom is at least that its dimension be greater than the wavelengths of transitions between the system's eigenstates. This might be the case for non-localized electrons in large molecules. Let us also mention here the case of strong interactions, in which transitions between states of the quark-antiquark system can have wave lengths of the same order as the extension of the states. Moreover, the coupling constant is not small. A calculation has been performed in (Billionnet 2004).

\bigskip

\centerline {\bf Appendix A. A third resonance, in the two level problem}
\smallskip

We must mention a third resonance. We are going to give its position when $\mu$ varies, $\lambda=0.1$ and $\delta=0.25$. We denote it by $\zeta_u(\mu)$, a function defined from its values for large $\mu$. More precisely, for $\mu=2$, $f_+(\kappa,\mu,\delta,.)$ has a zero at $\zeta_u=1.005-2.095\ i$, which differs from $\zeta_{w,ph}(2)$ which is $0.993-1.895\ i$. Following the displacement of this new resonance when $\mu$ decreases, we obtain $\zeta_u(\mu)$, represented by the curve in figure 10.
When $\mu$ tends to $0$, $\zeta_u$ tends to $1$ and when $\mu$ increases, $\zeta_u$ seems to be asymptotic to $1-i\bbbr^+$, behaving like $1-i\mu$. Now, for $\mu=2$, for instance, and $\lambda$ going to $0$, $\zeta_u$ tends to $1-i\mu$. We had the same behaviour for the nonstandard resonance $\zeta_{w,ph}(2)$, when $\lambda\rightarrow 0$.  
$$
\psannotate{\psboxto(5cm;0cm){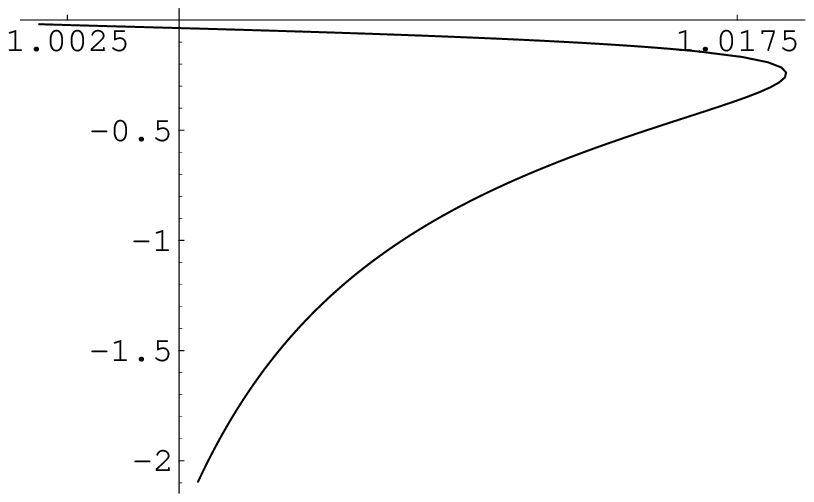}}{\at{1.5cm}{0cm}{$\leftarrow\mu=2$}\at{2.8cm}{2.5cm}{$^{\leftarrow}$$\mu$
  {\eightrm decreases}}}
$$
\medskip
\centerline {Figure 10. A third resonance. Variation with respect to $\mu$,
  for $\kappa=0.1$ and $\delta=0.25$}

\medskip

That there should be two zeros near $1-i\mu$ may be seen in the following way. Through deforming the integration contour in (8'), we can show that for $\Im
\zeta<0$,
$$
f_+(\kappa,\mu,\delta,\zeta)=f(\kappa,\mu,\delta,\zeta)-4i\ \kappa^2\
\mu^3\ \big({1\over (\mu^2+(\zeta+1)^2)^2}-{1\over (\mu^2+(\zeta-1)^2)^2}\big)\ .
$$
For $\Im \zeta<-\epsilon$ et $\Re \zeta>0$, we then have, for fixed $\mu$ and $\delta$,
$$
f_+(\kappa,\mu,\delta,\zeta)=\zeta-1-\delta-{c_2\
  \kappa^2\over (\zeta-(1-i\mu))^2}-{c_1\
  \kappa^2\over \zeta-(1-i\mu)}+o(\kappa^2)\ .
$$
For $\kappa$ small, one sees that, besides the zero near $1+\delta$, $f_+$ has indeed two zeros near  $\zeta=1-i\mu$, the limit of which is $1-i\mu$ when $\kappa$ goes to $0$.

The meaning of this resonance is mysterious. Whereas $\zeta_{w,ph}$ tended to the energy of the photon-atom mixed state $\zeta_{-,1}(\kappa,\delta)$ (here $\delta>0$) when the width of $g$ tended to $0$ (see 2.4.1 and (10a)), this new resonance seems to tend to $1$ (in units $\hbar c k_0$) whatever the coupling constant. The origin of $\zeta_u$ is the same as that of $\zeta_{w,ph}$, a pole of $g$. But whereas the physical interpretation of $\zeta_{w,ph}$ is quite clear from our study, we do not see any for $\zeta_u$ for the moment. There may also be none.

\bigskip

\centerline {\bf Appendix B. Calculation of two resonances for the hydrogen atom}
\smallskip
The radial parts of the wave functions of states  $\ket{n,n-1,0}$ and $ \ket{n-1,n-2,0}$ are
$$
R_1(r)=K_na_0^{-3/2}\ e^{-r/(na_0)} (r/a_0)^{n-1}\ ,\quad R_2(r)=K_{n-1}a_0^{-3/2}\ e^{-r/((n-1)a_0} (r/a_0)^{n-2}
$$
where $K_n=(2/n)^{n+1/2}\ ((2n)!)^{-1/2}$ normalizes the wave function. Let the subscript $n$ index all quantities related to the transition $\ket{n,n-1,0}\rightarrow \ket{n-1,n-2,0}$, with emission of a photon $(j=1,0,+1)$. The function  $g_{I,n}(k):=\obraket{1}{H_I}{2,\gamma}$ corresponding to this transition is (see for example (Moses 1973))
$$
g_{I,n}(y/\overline\rho)= -i{\sqrt 3\over 2\sqrt{\pi}}\alpha^{1/2}(e^2/\overline\rho_n)(\overline\rho_n/ a_0)^{2n}D_n\ \varphi_n(y)
$$
where
$$
D_n={K_nK_{n-1}\over n\sqrt{(2n-1)(2n-3)}},\quad \varphi_n(y):=y{\alpha_nQ_{2n-4}(y)+\beta_n(1+y^2)Q_{2n-5}(y)\over (1+y^2)^{2n-2}}
$$
with
$$
\alpha_n=2n^2-3n+2,\quad \beta_n=(2n-1)^2(n-2)
$$
$Q$ having been defined by (21).
$G$, $\mu$ and $\kappa$ in (20) are then, for the considered transition,
$$
G_n(y)=||\varphi_n||^{-1}\varphi_n(y)\ ,\quad\quad \mu_n=n(n-1)\ E_I^{-1}\hbar c/a_0
$$
and
$$
\kappa_n={(3\alpha)^{1/2}\over 2\sqrt{\pi}}(e^2/\overline\rho_n)(\overline\rho_n/ a_0)^{2n}D_n\ {\cal E}_{n,n-1}^{-1}||\varphi_n||={(3\alpha)^{1/2}\over 2\sqrt{\pi}}{e^2/a_0\over E_I}\ {(n(n-1))^{2n+1}\over (2n-1)^{2n}}\ D_n\ ||\varphi_n||\ .
$$
Through using $e^2/a_0\simeq 2E_I$, we get $\kappa_n\simeq \sqrt{3/\pi}\ \alpha^{1/2}\ (n(n-1))^{2n+1} (2n-1)^{-2n}\ D_n\ ||\varphi_n||$.
A computer gives
$$
\kappa_2=0.018,\quad \kappa_{10}=0.022,\quad \kappa_{50}=0.028
$$
and, for $n=2$, 
$$
\mu_2\simeq548\quad{\rm and}\quad G_2(y)=-\sqrt 3\ {y\over (1+y^2)^2} \ .
$$
The two zeros of F are then respectively $\zeta_{2,{\rm s}}\simeq 1-2\times 10^{-6}-2\times 10^{-8}i$ and $\zeta_{2,{\rm n.s.}}=1.493-544\ i$.

\bigskip

\centerline{\bf References}

\medskip

\item{}
Arai A 1989 {\it J. Math. Anal. Appl.} {\bf 140} 270

\smallskip
\item{} Billionnet C 2001 {\it J. Phys.} A {\bf 34} 7757

\smallskip
\item{} Billionnet C 2002 {\it J. Phys.} A {\bf 35} 2649

\smallskip
\item{} Billionnet C 2004 {\it Int. J. Mod. Phys.} {\bf 19} 2643

\smallskip
\item{} Billionnet C 2005 {\it J. Math. Phys.} {\bf 46} 072101 

\smallskip
\item{}
Boeuf F, Andr\'e R, Romestain R, Le Si Dang, P\'eronne E, Lampin J F, Hulin D and  Alexandrou A 2000 {\it Phys. Rev.} B {\bf 62} R2279

\smallskip
\item{} Callaway J 1974 {\it Quantum Theory of the Solid State} (New York: Academic Press)

\smallskip
\item{}
Cohen-Tannoudji C, Diu B and Lalo\"e F 1973 {\it M\'ecanique Quantique} (Paris: Hermann)

\smallskip
\item{}
Cohen-Tannoudji C, Dupont-Roc J and Grynberg G 1988 {\it Processus d'interaction entre photons et atomes} (Paris: CNRS) (Engl. transl. 1992 (New York: Wiley)

\smallskip
\item{}
Hameau S, Guldner Y, Verzelen O, Ferreira R, Bastard G, Zeman J, Lema\^\i{}tre A and G\'erard J M 1999 {\it Phys. Rev. Lett.} {\bf 83} 4152

\smallskip
\item {} Haroche S 1984, {\it Rydberg atoms and radiation in a resonant cavioty}, in Les Houches, XXXVIII, (Amsterdam: North Holland) 

\smallskip
\item {} Haroche S 1992, {\it Cavity Quantum Electrodynamics}, in Les Houches, LIII, North Holland 

\smallskip
\item {} Haroche S and  Raimond J M 1993, {\it Sci. Am.} {\bf 268} (4) 26

\smallskip
\item{}
Inoshita T and Sakaki H 1997 {\it Phys. Rev.} B {\bf 56} R4355

\smallskip
\item{} Moses 1973 {\it Phys. Rev.} A {\bf 8} 1710

\smallskip
\item {} Raimond J-M, Brune M and Haroche S 2001 {\it Rev. Mod. Phys.}, {\bf 73} 565

\smallskip
\item{}
Reed M. 1993 {\it Sci. Am.} {\bf 268} (1) 98

\smallskip
\item {} Senellart P 2003 {\it Ann. Phys.} {\bf 28} (4).

\smallskip
\item{}
Sermage B, Long S, Abram I, Marzin J Y, Bloch J, Planel R and Thierry-Mieg V 1996 {\it Phys. Rev.} B {\bf 53} 16516

\smallskip
\item{}
Tignon J, Voisin P, Delalande C, Voos M, Houdr\'e R, Oesterle U and Stanley R P  1995 {\it Phys. Rev. Lett.} {\bf 74} 3967

\smallskip
\item{}
Verzelen O, Ferreira R and Bastard G 2000 {\it Phys. Rev.} {\bf 62} B  R4809

\smallskip
\item{}
Verzelen O, Ferreira R and Bastard G 2002, {\it Phys. Rev. Lett.} {\bf 88} 146803

\smallskip
\item{}
Weisbuch C, Nishioka M, Ishikawa A, Arakawa Y 1992 {\it
    Phys. Rev. Lett.} {\bf 69} 3314

\smallskip
\item{}Weisbuch C and Vinter B 1991 {\it Quantum Semiconductor Structures} (San Diego: Academic Press)

\smallskip
\item {} Yamamoto Y, Tassone F and Cao H 2000 {\it Semiconductor Cavity Quantum Electrodynamics} (Berlin: Springer)

\end